\documentclass[timestamp,acmtog]{acmart}

\setcopyright{rightsretained}
\copyrightyear{2023}
\acmYear{2023}
\acmDOI{XXXXXXX}

\acmJournal{TOG}
\acmVolume{X}
\acmNumber{X}
\acmArticle{X}
\acmYear{2023}
\acmMonth{4}

\usepackage{booktabs}

\graphicspath{{./images/}}

\usepackage{wrapfig}

\usepackage{comment}

\usepackage[linesnumbered]{algorithm2e}

\usepackage{xargs} 

\newif\ifsubmit
\submittrue
\ifsubmit
    \newcommand{\todo}[2][1=]{}
    \newcommand{\TODO}[1]{}
    
    \newcommand{\yotam}[2][1=]{}
    \newcommand{\ted}[2][1=]{}
    \newcommand{\jason}[2][1=]{}
    \newcommand{\jianchao}[2][1=]{}
    \newcommand{\jose}[2][1=]{}

    \newcommand{\topic}[1]{}

    \newcommand{\revbegin}{}
    \newcommand{\revend}{}
\else
    \definecolor{todo}{rgb}{0.99,0.59,0.55}
    \newcommand{\TODO}[1]{\textcolor{red}{\textbf{****** #1 ******}}}
    
    \definecolor{cyan3}{HTML}{C99E10} %
    \usepackage[colorinlistoftodos,prependcaption,textsize=tiny,color=todo,tickmarkheight=0.1cm]{todonotes}
    \setuptodonotes{noinline}
    
    \definecolor{lime}{rgb}{0.53,0.78,0.27}
    \newcommandx{\yotam}[2][1=]{\todo[linecolor=lime,backgroundcolor=lime!25,bordercolor=lime,#1]{\colorbox{lime!50}{\textsc{Yotam:}} #2}}
    \newcommandx{\ted}[2][1=]{\todo[linecolor=teal,backgroundcolor=teal!25,bordercolor=teal,#1]{\colorbox{teal!50}{\textsc{Ted:}} #2}}
    \newcommandx{\jason}[2][1=]{\todo[linecolor=orange,backgroundcolor=orange!25,bordercolor=orange,#1]{\colorbox{orange!50}{\textsc{Jason:}} #2}}
    \newcommandx{\jianchao}[2][1=]{\todo[linecolor=blue,backgroundcolor=blue!25,bordercolor=blue,#1]{\colorbox{blue!50}{\textsc{Jianchao:}} #2}}
    \newcommandx{\jose}[2][1=]{\todo[linecolor=magenta,backgroundcolor=magenta!25,bordercolor=magenta,#1]{\colorbox{magenta!50}{\textsc{Jose:}} #2}}
    
    \newcommand{\revbegin}[1]{\color{blue}} 
    \newcommand{\revend}[1]{\color{black}} 

    \usepackage[plain]{markdown}
\fi

\usepackage{xspace}

\begin{document}


\title{Text-guided Image-and-Shape Editing and Generation: A Short Survey}


\author{Cheng-Kang Ted Chao}
\email{cchao8@gmu.edu}
\affiliation{%
  \institution{George Mason University}
  \country{USA}
}

\author{Yotam Gingold}
\email{ygingold@gmu.edu}
\affiliation{%
  \institution{George Mason University}
  \country{USA}
}

\renewcommand{\shortauthors}{}

\begin{abstract}
Image and shape editing are ubiquitous among digital artworks. Graphics algorithms facilitate artists and designers to achieve desired editing intents without going through manually tedious retouching. In the recent advance of machine learning, artists' editing intents can even be driven by \emph{text}, using a variety of well-trained neural networks. They have seen to be receiving an extensive success on such as generating photorealistic images, artworks and human poses, stylizing meshes from text, or auto-completion given image and shape priors. In this short survey, we provide a overview over 50 papers on state-of-the-art (text-guided) image-and-shape generation techniques. We start with an overview on recent editing algorithms in the introduction. Then, we provide a comprehensive review on text-guided editing techniques for 2D and 3D independently, where each of its sub-section begins with a brief background introduction. We also contextualize editing algorithms under recent implicit neural representations. Finally, we conclude the survey with the discussion over existing methods and potential research ideas.
\end{abstract}




\maketitle

\section{Introduction}
Image editing tasks are large research area in computer graphics.
Over the past forty years, a variety of image editing algorithms are proposed to enable a new level of creativity for digital raster or vector arts. There are two types of image editing algorithms. The first type is automatic approaches, that is, generating outputs directly based on different editing purposes. For example, image stylization \cite{hertzmann2001image, li2018closed, li2019learning}, text-guided image generation \cite{ramesh2021zero, nichol2021glide} and abstraction \cite{vinker2022clipasso}.
This type of algorithms is typically not tailored to allow users to \emph{control} over output images after the algorithms are applied. The other type is semi-automatic approaches. The core of this type of algorithms is to design \emph{handles} for users to manipulate image contents. To provide different editing purposes for image manipulations, \emph{handles} are designed/computed as, e.g. palette colors (simplified convex hull) for image recoloring \cite{tan2018efficient, Tan:RGB2016, wang2019improved, Chang:2015:PPR}, color blends \cite{chao2021posterchild} and sketching strokes \cite{ha2017neural} for image abstraction, or soft layers \cite{aksoy2017unmixing, aksoy2018semantic, koyama2018decomposing} for color/semantic editing. More recently, \emph{text} has also been treated as \emph{handles} to edit image features \cite{patashnik2021styleclip, kim2022diffusionclip, avrahami2022blended}. Though most algorithms are designed for raster images (due to its natural grid-based properties that favors convolutions in neural networks), some recent works have been seen to bridge the gap between raster and vector graphics \cite{li2020differentiable, reddy2021im2vec, bessmeltsev2019vectorization}. 

Additional degree of freedom on 3D shapes allows more flexible editing possibilities but also yields more complexities. Here, we briefly introduce editing algorithms under representations on point clouds, voxels and meshes. Specifically, we categorize these algorithms into two parts, stylization and generation. For stylization, recent techniques include cubic stylization \cite{liu2019cubic}, where they formulate a as-rigid-as-possible energy with $l_1$-regularization to create stylized geometries. Gauss stylization \cite{kohlbrenner2021gauss} generalizes \cite{liu2019cubic}'s work to using surface normals for more flexible and interactive control. Adversarial approaches are proposed by \cite{hertz2020deep} to synthesize geometric textures on given meshes. Textures on meshes can also be stylized by leveraging text \cite{michel2022text2mesh}. Without direct involvement from human editing intents, shape generation has been proposed by using point-voxel diffusion \cite{zhou20213d}, generative adversarial networks (GANs) \cite{ibing20213d}, and DSG-Net \cite{yang2022dsg}.

While above discrete representations provide an easy way for computations, the memory cost of parametrization over finer resolution become a problem. Recent implicit neural representation (INR) overcomes this shortcoming by encoding implicit function values, e.g. signed distance \cite{park2019deepsdf}, colors \cite{dupont2021coin} and image gradients \cite{sitzmann2020implicit}, or occupany fields \cite{mescheder2019occupancy}, using a multi-level perceptron (MLP) with a fixed number of parameters. This continuous representation yields better memory-efficiency and theoretically infinite resolution. However, direct editibility under implicit representations remains an open problem since network weights are normally unexplainable. To mitigate this issue, recent work uses auxiliary latent code as handles to achieve smooth shape interpolation with Lipschitz regularizer \cite{liu2022learning}. \citet{mehta2022level} uses level-set theory to allow shape deformation operations applied on implicit surfaces. 

\begin{figure}
\includegraphics[width=0.9\linewidth]{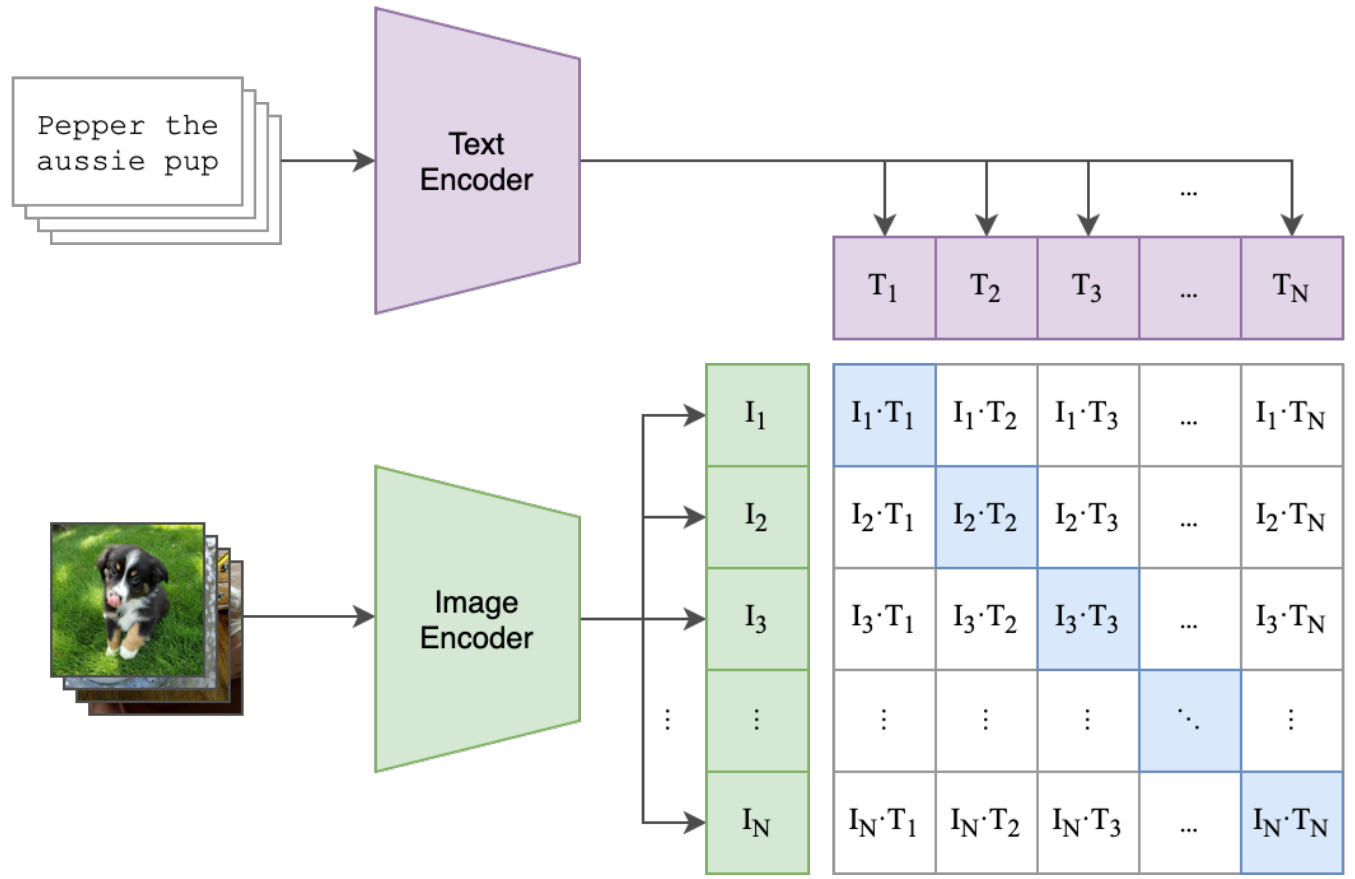}
\caption{High-level idea of how CLIP is trained. CLIP trains jointly on text and image encoders, which yields a multi-modal embedding space. CLIP can be used for zero-shot prediction, image synthesis \cite{patashnik2021styleclip, abdal2022clip2stylegan} and image abstraction \cite{vinker2022clipasso}. (\emph{image is from original CLIP paper \cite{radford2021learning}}.)}
\label{fig:clip}
\end{figure}

Above methods give a short overview of current editing algorithms in recent years. In the following of this survey, we go through the typical techniques used in state-of-the-art editing and generative algorithms, specifically and mostly, using \emph{text} as a guidance. We begin with a background introduction on Contrastive Language-Image Pre-training (CLIP) \cite{radford2021learning}, a large vision-language model. Then, we review the techniques using CLIP as a base model to perform zero-shot or few-shot learning using e.g. diffusion model \cite{sohl2015deep, ho2020denoising}, GANs, or variational auto-encoder (VAEs) \cite{kingma2013auto}, for image-and-shape editing and generation. We also briefly include some techniques using other means as guidance, e.g. edge map, masks, segmentations. Finally, we discuss current editing techniques under INR and we conclude the survey with possible future editing directions.  

\subsection{Related Surveys}
For editing and synthesis, there are survey papers focusing specifically on GANs \cite{wu2017survey, liu2022survey}, faces using StyleGAN \cite{melnik2022face} and multimodal guidances \cite{zhan2021multimodal}, as well as shape generation \cite{xu2022survey}, mesh segmentation \cite{shamir2008survey}, and shape editing \cite{yuan2021revisit}. Though our survey review is relatively short, we share the same motivation as these surveys, aiming to provide consolidate knowledge over methods in recent years among graphics communities.

\section{Background: Text-guided Editing}
Leveraging natural language descriptions as input to provide editing can be traced back to early 80's, where \citet{adorni1984natural} built a system to imagine static scene from a sequence of simple phrases. This motivates many following works, including WordsEye \cite{coyne2001wordseye}, which extends Put system \cite{clay1996put} to allowing more flexible arrangements of 3D objects to render 3D scene from sentences. More recently, a large joint vision-language model, CLIP, was proposed by OpenAI \cite{radford2021learning}. CLIP is trained on 400 million text-image pairs from common-crawl dataset with contrastive objectives to learn image representations from text. The overall framework can be seen in Figure~\ref{fig:clip}. 

To be more specific, the image encoder is usually chosen as ResNet \cite{he2016deep} or Vision Transformer (ViT) \cite{dosovitskiy2020image}, and the text encoder is a Transformer \cite{vaswani2017attention}. During training, given a batch of $N$ text-image pairs (which yields $N \times N$ possible pairs), CLIP learns two linear projections from image and text embedding space, respectively, to multi-modal embedding space. CLIP is trained to maximize the cosine similarity of $N$ correct pairs and minimize the cosine similarity of $N^2 - N$ incorrect pairs with symmetric cross entropy loss.  

\begin{figure}
\includegraphics[width=\linewidth]{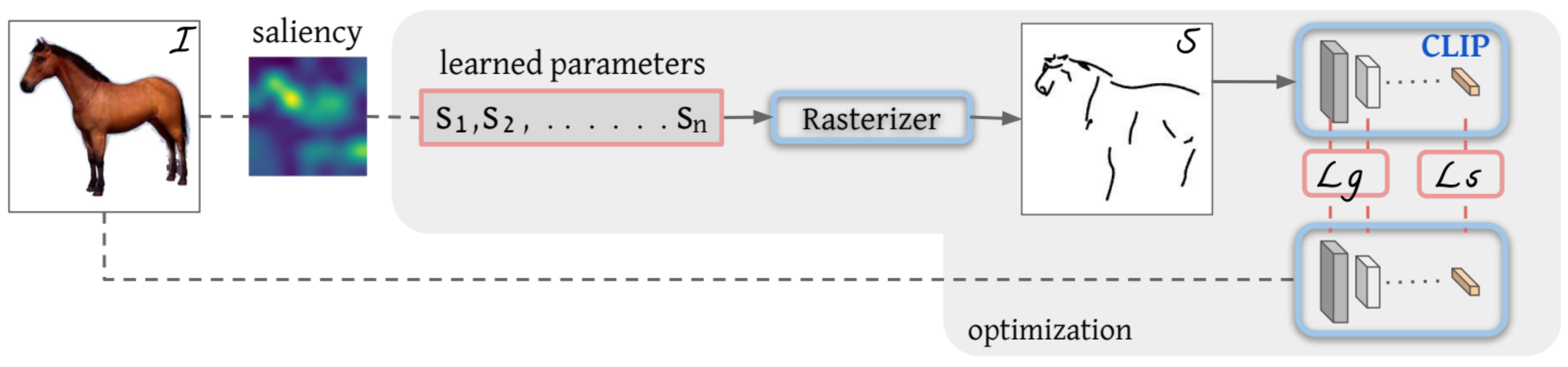}
\caption{CLIPasso \cite{vinker2022clipasso} first extracts a saliency map from given image $I$ to sample stroke locations $\{S_1, S_2, \cdots, S_n\}$. The locations are then passed through a differentiable rasterizer to obtain a rasterized rough sketch. The parameters (i.e. locations) are optimized through back-propagation via the CLIP losses. (\emph{figure is from original CLIPasso paper \cite{vinker2022clipasso}}.)}
\label{fig:clipasso}
\end{figure}

The capabilities of CLIP open up many possible applications for zero-shot learning editing. For instance, CLIPasso \cite{vinker2022clipasso} employs a differentiable rasterizer on stroke-based vector graphics defined by a set of bezier control points, and encode both original image and rasterized image into CLIP embedding space to compare the CLIP loss (also along with some geometric loss in certain layers of CLIP image encoder). Then, the locations of control points are then optimized through back-propogation to achieve semantically-aware sketch-based image abstraction (Figure~\ref{fig:clipasso}). We provide more details of other applications in the following sections.

\section{Method}
We divide the problems into image and shape domains. However, many of these algorithms heavily build upon framework from GANs, VAEs and diffusion models. Therefore, we start the conversation over high-level and brief introduction of these models. Then, we provide several state-of-the-art techniques on how the above models can be used to drive editing in each domain. Finally, we conclude each subsection with discussion over strengths and weaknesses of these methods.

\paragraph{Generative Adversarial Networks (GANs)} A part of unsupervised learning is \emph{generative modeling}, in which it is very useful when human-annotated data are limited. Traditional approaches explicitly model the data distribution by a set of parameters. Then, it finds the best parameters to maximize objective functions, i.e. maximum likelihood estimation. Unlike explicit modeling, GANs, first proposed by \cite{goodfellow2014generative}, is an implicit approach that employs a neural network, i.e. a \emph{generator} $G$, to capture data distribution. GANs has another neural network, i.e. a \emph{discriminator} $D$, examines samples generated from $G$ and produces an estimation of whether the generated output is real or not. These two neural networks are trained jointly, which mimics a minimax two-player game under this adversarial framework. Mathematically, the goal of training a GANs is to find a \emph{local Nash equilibrium}, i.e. according to \cite{goodfellow2020generative}, by minimizing
\begin{equation} \label{eq:gan_loss}
\begin{aligned}
J^{(G)}\left( \theta^{(G)}, \arg\min_{\theta^{(D)}} J^{(D)}( \theta^{(G)}, \theta^{(D)}) \right)
\end{aligned}
\end{equation}
where $\theta^{(G)}$ is generator's learnable parameters and $J^{(G)}$ is the loss for $G$. The same applies \emph{mutatis mutandis} to discriminator $D$. In other words, each neural network tries to minimize its cost. In practice, GANs is commonly trained using gradient-based optimization on $G$ and $D$, alternatively. Despite GANs exhibiting the capacity to produce realistic and high-quality images, GANs is notoriously difficult to train due to the adversarial nature of the training process. GANs requires $G$ and $D$ to compete with each other, which can lead to "mode collapse", i.e. $G$ finds a way to generate only a limited ranges of outputs to mislead $D$. GANs is also sensitive to hyperparameters. However, recent advances in GANs architecture and training techniques have shown promise in addressing these issues and improving the stability and performance of GANs \cite{karras2017progressive, zhang2019self}. 

\paragraph{Variational Autoencoders (VAEs)} Generative process using a decoder from simple trained autoencoders (AEs) is extremely difficult since its latent space is normally not "well-organized". Contrary to encoder in AEs (encoding an input to a single point in the latent space), VAEs \cite{kingma2013auto} encodes an input as a distribution. VAEs is then trained along with a regularization term (Eq.~\ref{eq:vae_loss}), i.e. Kullback-Leibler (KL) divergence, to provide exploitable latent space. Formally, given $\mathbf{x}$ representing the input data, and $\mathbf{z}$ representing the compressed latent space representation. The encoder $h_\phi$ encodes input $\mathbf{x}$ and outputs a distribution over the latent space (in practice, maps $\mathbf{x}$ to $\mathbf{\mu_x}$ and $\mathbf{\sigma_x}$ to approximate posterior distribution as Gaussian). We denote $f_\theta$ as the decoder. Therefore, the loss for training a VAEs is to minimize
\begin{equation} \label{eq:vae_loss}
\begin{aligned}
\mathcal{L}_{\text{VAE}} = \underbrace{\mathbb{E}_{z \sim q_{\phi}(\mathbf{z}\mid\mathbf{x})}[-\log p_{\theta}(\mathbf{x}\mid\mathbf{z})]}_{\text{Reconstruction loss}} + \underbrace{\text{KL}[q_{\phi}(\mathbf{z}\mid\mathbf{x})\parallel p(\mathbf{z})]}_{\text{Regularization term}}
\end{aligned}
\end{equation}
with 
\begin{equation} \label{eq:vae_posterior}
\begin{aligned}
 h_\phi(x) = \mu_x, \hspace{1mm} \sigma_x\\
q_{\phi}(\mathbf{z}\mid\mathbf{x}) = \mathcal{N}(\mu_x, \sigma_x)\\
p_{\theta}(\mathbf{x}\mid\mathbf{z}) = e^{\frac{-\|\mathbf{x} - f(\mathbf{x})\|^2}{2c}}
\end{aligned}
\end{equation}
where $p(\mathbf{z})$ is often modeled as a normal distribution with zero mean and identity covariance matrix, i.e. $\mathcal{N}(0, I)$. The first term in Eq.~\ref{eq:vae_loss} refers to minimizing the expected negative log-likelihood of $\mathbf{x}$ given $\mathbf{z}$ sampled from $q_{\phi}(\mathbf{z}\mid\mathbf{x})$. The second term can be seen as a regularization over $\phi$ to approximate posterior distribution to normal distribution. Note that in practice, $\mathbf{z}$ is computed as $\mathbf{z} = \sigma_x \cdot \xi + \mu_x$ with $\xi \sim \mathcal{N}(0, I)$, to allow back-propagation for gradients. Though VAEs has proved to learn a regular latent space, a major drawback is \emph{posterior collapse}, where the learned latent representation of the data becomes uninformative, i.e. collapsing to prior normal distribution $p(\mathbf{z})$, resulting in poor generalization and inability to generate diverse samples. Some remedies include, e.g. beta-VAE \cite{higgins2017beta, burgess2018understanding}, as well as vector-quantized VAE \cite{van2017neural}.

\begin{figure} 
\includegraphics[width=\linewidth]{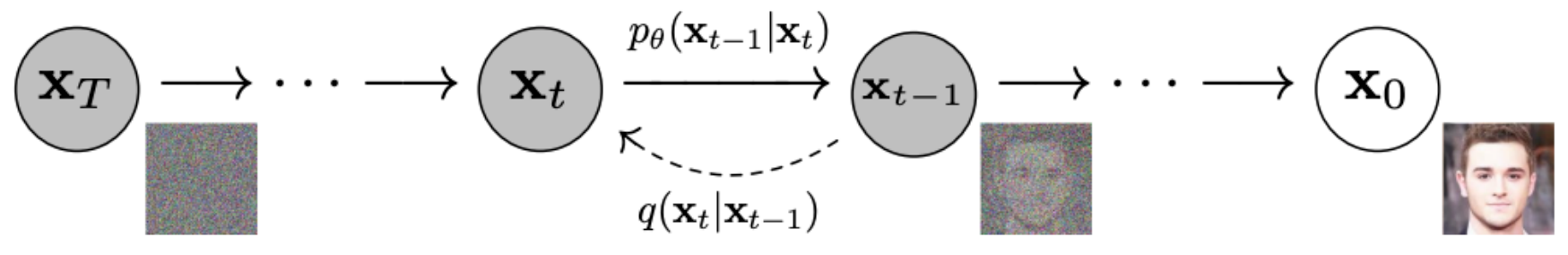}
\caption{Illustration of framework for denoising diffusion probabilistic model. The forward process is denoted as $q(\mathbf{x}_t\mid\mathbf{x}_{t-1})$. A neural network is used to predict noise $\epsilon_\theta(\mathbf{x}_{t}, t)$ which makes $\mathbf{x}_{t-1}$ into $\mathbf{x}_{t}$. In practice, U-Net \cite{ronneberger2015u} is often used as a backbone for the reverse process. (\emph{image source: \cite{ho2020denoising}}.)}
\label{fig:ddpm}
\end{figure}

\paragraph{Denoising Diffusion Probabilistic Models (DDPM)} \label{paragraph:ddpm} Diffusion probabilistic models, inspired from non-equilibrium statistical physics \cite{sohl2015deep}, are proposed to introduce random (Gaussian) noise to data in a gradual manner, behaving a process of Markov chain (Figure~\ref{fig:ddpm}). Theoretically, under large number of steps of this forward pass, the data approximate pure Gaussian noise. Then, the reverse of this process, i.e. denoising, is learned by a neural network. The network then uses the predicted noise to reconstruct the original data from the noisy input, which yields a generative process. This method has been popularized by DDPM \cite{ho2020denoising}, performing high-quality image synthesis, and has shown superior performance on image generation tasks among state-of-the-art generative models \cite{dhariwal2021diffusion}. From a probabilistic standpoint, given a data point $\mathbf{x}_0$ sampled from data distribution $q(\mathbf{x}_0)$, the forward process is defined as
\begin{equation} \label{eq:dm_forward_process}
\begin{aligned}
q(\mathbf{x}_t\mid\mathbf{x}_{t-1}) \coloneqq \mathcal{N}(\sqrt{1 - \beta_t} \mathbf{x}_{t-1}, \beta_t \mathbf{I})
\end{aligned}
\end{equation}
where $\beta_t$ is some constant and $\mathbf{x}_{t}$ is a variable for $t = 1, \cdots, T$. We denote the reverse process as $p_\theta(\mathbf{x}_{t-1}\mid\mathbf{x}_{t})$ (which is learned by a neural network), i.e.
\begin{equation} \label{eq:dm_reverse_process}
\begin{aligned}
p_\theta(\mathbf{x}_{t-1}\mid\mathbf{x}_{t}) \coloneqq \mathcal{N}(\mathbf{x}_{t-1};  \mu_\theta(\mathbf{x}_{t}, t), \sigma_t^2 \mathbf{I})
\end{aligned}
\end{equation}
Note that mathematically, $\mu_\theta(\mathbf{x}_{t}, t)$ (in Eq.~\ref{eq:dm_reverse_process}) can be substituted as a function of $\mathbf{x}_{t}$ and $\epsilon_\theta(\mathbf{x}_{t}, t)$ (noise predicted by a network with parameters $\theta$). Therefore, according to \cite{ho2020denoising}, the training can be simplified as approximating $\epsilon_\theta(\mathbf{x}_{t}, t)$ to ground truth noise $\epsilon \sim \mathcal{N}(0, \mathbf{I})$ by minimizing
\begin{equation} \label{eq:dm_loss}
\begin{aligned}
\mathcal{L}_{\text{ddpm}} = \mathbb{E}_{t\sim [1,T], \mathbf{x}_0, \epsilon} \left[ \| \epsilon - \epsilon_\theta(\mathbf{x}_{t}, t) \|^2 \right]
\end{aligned}
\end{equation}
In addition, $\mathbf{x}_t$ can be reparameterized as 
\begin{equation} \label{eq:dm_x_eps}
\begin{aligned}
\mathbf{x}_{t} = \sqrt{\Bar{\alpha_t}} \mathbf{x}_{0} + \sqrt{1 - \Bar{\alpha_t}} \epsilon
\end{aligned}
\end{equation}
for constant $\Bar{\alpha_t}$. Therefore, the training algorithm (modified from \cite{ho2020denoising}) can be concluded in Algorithm~\ref{alg:ddpm_train}. The sampling, i.e. denoising process, can be implemented as in Algorithm~\ref{alg:ddpm_sample} by simplifying variational lower bound and loss reparametrization \cite{ho2020denoising}. This is done by (1) starting with a random noise and a timestamp $t = T$ (2) subtracting noise predicted by the networks with linear scheduling (3) looping through step 2 until $t = 1$.

\RestyleAlgo{ruled}
\begin{algorithm}[hbt!]
\SetKwComment{Comment}{/* }{ */}
\SetKwFunction{Rep}{Repeat}
\SetKwFunction{Uni}{Uniform}
\caption{DDPM Training}\label{alg:ddpm_train}
\While{converged}{
$\mathbf{x}_0 \sim q(\mathbf{x}_0)$  \Comment{sample a data point}
$t \sim \Uni(\{1, \cdots, T\})$ \Comment{sample a timestamp}
$\mathbf{\epsilon} \sim \mathcal{N}(0, \mathbf{I})$ \Comment{compute random noise}
Take gradient descent step on
\hspace{3mm} $\nabla_\theta \left[\| \epsilon - \epsilon_\theta(\sqrt{\Bar{\alpha_t}} \mathbf{x}_{0} + \sqrt{1 - \Bar{\alpha_t}} \epsilon, t) \|^2 \right]$
}
\end{algorithm}

\vspace{-6mm}
\RestyleAlgo{ruled}
\begin{algorithm}[hbt!]
\SetKwComment{Comment}{/* }{ */}
\SetKwFunction{Given}{Given}
\SetKwFunction{Return}{Return}
\SetKwFunction{Uni}{Uniform}
\caption{DDPM Sampling}\label{alg:ddpm_sample}
\Given: $T$ \Comment{given a timestamp}
$\mathbf{x}_T \sim \mathcal{N}(0, \mathbf{I})$ \Comment{sample random noise}
\For{t = T, T-1, ..., 1}{
$\mathbf{z} \sim \mathcal{N}(0, \mathbf{I})$ if $t > 1$; else $\mathbf{z} = 0$\\
$\mathbf{x}_{t-1} = \frac{1}{\sqrt{\alpha_t}}(\mathbf{x}_t - \frac{1 - \alpha_t}{\sqrt{1 - \Bar{\alpha}}_t} \epsilon_\theta(\mathbf{x}_t, t)) + \sigma_t \mathbf{z}$
}
\Return $\mathbf{x}_0$
\end{algorithm}

\subsection{Image Problems} \label{sec:image_problems}
Learning-based image editing algorithms, which are trained on large datasets of images, have shown powerful ability to generate images with high precision and fidelity. However, the editability of the generative process may be limited, as the algorithms are often designed to learn a latent space by optimizing a specific objective function, which may not always align with the user's desired editing intents. In this section, we discuss state-of-the-art techniques that address these limitations. These approaches incorporate \emph{text} (and also some other guidances) into the learning process, commonly along with using CLIP, creating more interpretable models for user control. We categorize these techniques into GAN-based (Sec.~\ref{sec:gan-based}), diffusion-based (Sec.~\ref{sec:diffusion-based}) and zero-shot (Sec.~\ref{sec:zero-shot}) approaches. 

\begin{figure} 
\includegraphics[width=\linewidth]{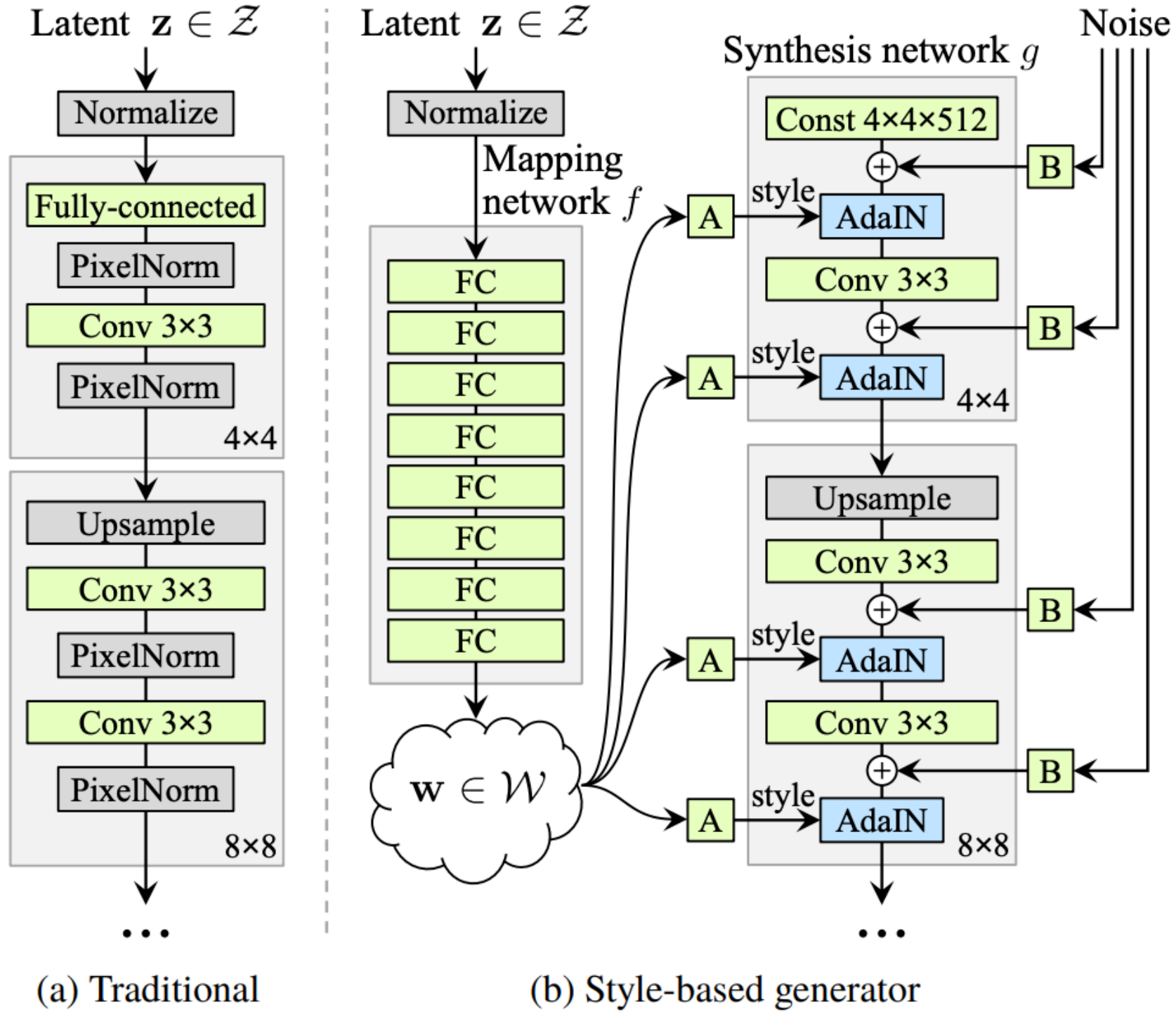}
\caption{StyleGAN framework. Traditional generator, i.e. ProGAN \cite{karras2017progressive}, provides a stable strategy for training a generator by progressively increasing image resolutions (e.g. from 4x4, 8x8, ..., to 1024x1024). StyleGAN \cite{karras2019style} extends this idea by introducing latent code $w$ from a learned intermediate latent space $\mathcal{W}$ to coarse-to-fine resolution layers in a synthesis network $g$. (\emph{image source: \cite{karras2019style}}.)}
\label{fig:stylegan}
\end{figure}

\subsubsection{GAN-based Approaches} \label{sec:gan-based}
Generators have been a popular framework for performing image synthesis. A notable recent breakthrough on image editing using GANs is StyleGAN, proposed by \cite{karras2019style} (and also some of its variants, a.k.a. StyleGAN2 \cite{karras2020analyzing}, StyleGAN3 \cite{karras2021alias}). StyleGAN has shown powerful ability to generate and interpolate high-quality and realistic human portraits. It extends the idea of ProGAN \cite{karras2017progressive} by feeding a latent code $w$ from a intermediate latent space $\mathcal{W}$ into a synthesis network at different upsampling layers to control coarse-to-fine semantic details (Figure~\ref{fig:stylegan}). 

\begin{figure} 
\includegraphics[width=\linewidth]{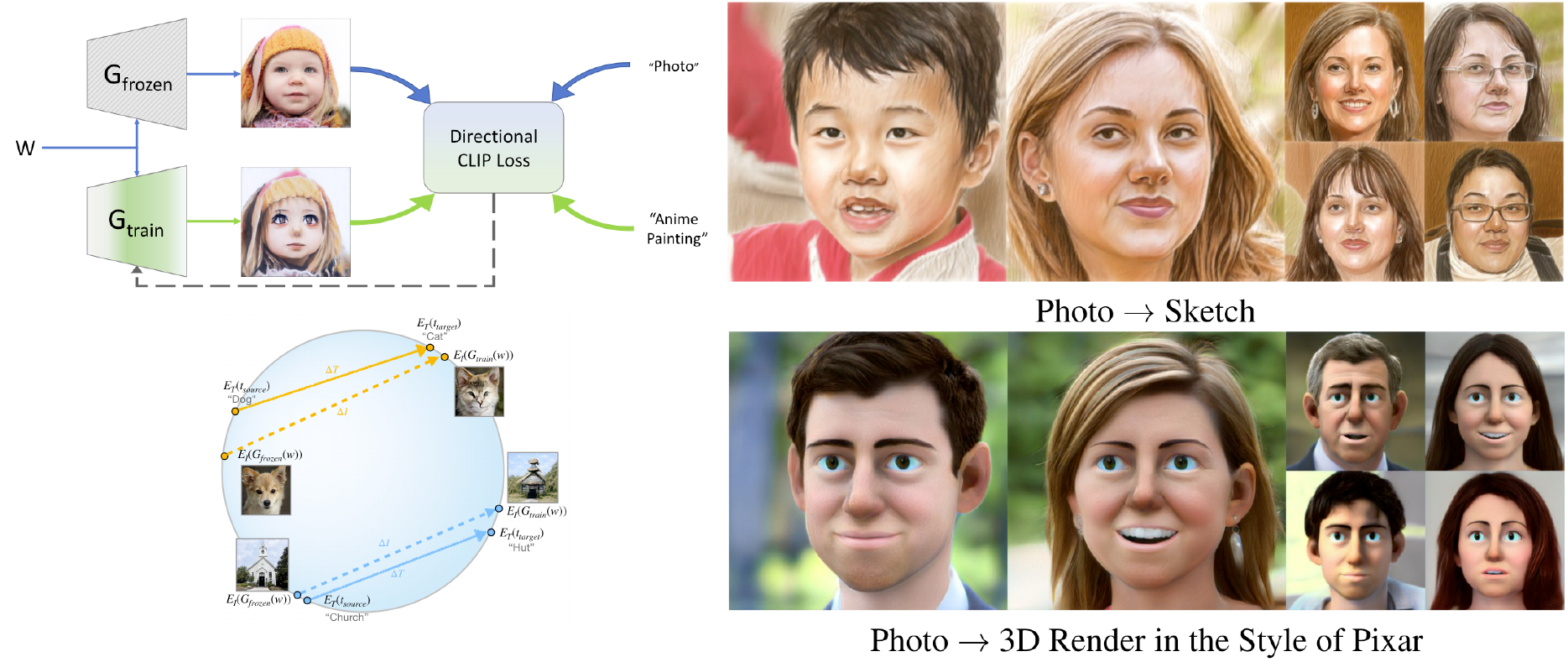}
\caption{Given a latent code $w$, input text prompt "Photo" and target text prompt "Anime Painting" (top left), StyleGAN-NADA computes directional CLIP loss in CLIP space (bottom left) from generated images returned from a frozen generator and a trained generator. This allows the optimization to be regularized to avoid adversarial solutions, yielding domain shifting using text (right).  (\emph{images cropped from: \cite{gal2021stylegan}}.)}
\label{fig:stylegan-nada}
\end{figure}

\begin{figure} 
\includegraphics[width=\linewidth]{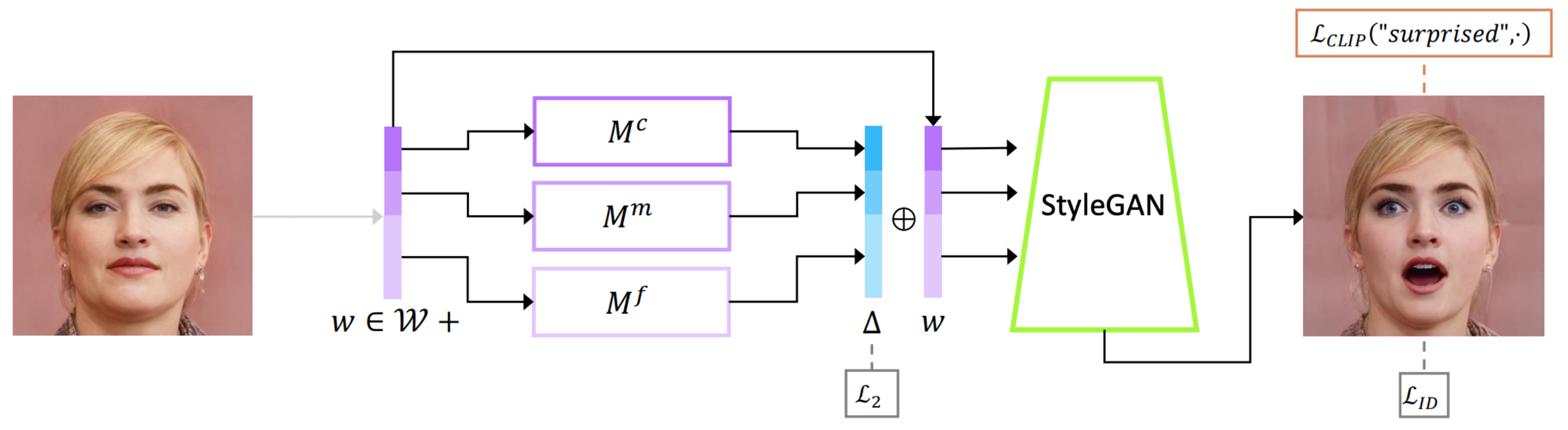}
\caption{The input image (left) is first embedded into a latent space with latent code $w$. StyleCLIP then uses three separate learnable mapping functions to generate residuals $\Delta$ (in blue), which is used to be added with $w$. The new latent code is then fed into a pre-trained StyleGAN (in green) to generate images (right) reflecting given text prompt (in this case "\emph{surprised}"). This optimization is evaluated by CLIP loss and identity loss. (\emph{image source: \cite{patashnik2021styleclip}}.)}
\label{fig:styleclip}
\end{figure}

Many of GAN-based approaches build upon StyleGAN. For example, StyleCLIP \cite{patashnik2021styleclip} performs text-driven image manipulations under StyleGAN's latent space. This is done by optimization over a latent code in StyleGAN's $\mathcal{W}^+$ space \footnote{$\mathcal{W}^+$ space is an extended StyleGAN's latent space proposed by \cite{abdal2019image2stylegan}. Note that StyleGAN aims for realistic generation but not for editing over a given image. \citet{abdal2019image2stylegan}'s approach allows any given image to be embedded into the latent space for semantic manipulation under pre-trained StyleGAN.} by minimizing loss from CLIP space along with identity loss (Figure~\ref{fig:styleclip}). The CLIP loss is measured by cosine distance between the embedding of generated image and text prompt embedding in CLIP space. The identity loss is defined from ArcFace \cite{deng2019arcface}. StyleCLIP further inspires many following works. These works all share the same motivation: \emph{how to find semantic editing directions in CLIP space?} To address this problem, StyleGAN-NADA \cite{gal2021stylegan} uses a generator as a regularizer to shift the domain of pre-trained generator to another domain using text by minimizing a directional CLIP loss (Figure~\ref{fig:stylegan-nada}). \citet{abdal2022clip2stylegan} finds semantic editing directions (from a large set of images within same category, e.g. human faces) by a disentanglement optimization in CLIP space and perform text-guided semantic editing using extracted directions with StyleGAN generator. 

GANs can also be trained with conditioning on text to achieve text-to-image generation  \cite{reed2016generative, zhang2017stackgan, zhang2018stackgan++, qiao2019learn}, i.e. generate an image reflecting given text description. Specifically, recent state-of-the-art methods include, Df-GAN \cite{tao2020df}, which is a simple yet effective model consisting of one-stage text-to-image framework. They incorporate novel matching-aware gradient penalty and one-way output in the discriminator to maintain semantic consistency. MirrorGAN \cite{qiao2019mirrorgan} trains end-to-end under text-to-image-to-text framework to guarantee strong text semantic alignment. 

\begin{figure} 
\includegraphics[width=\linewidth]{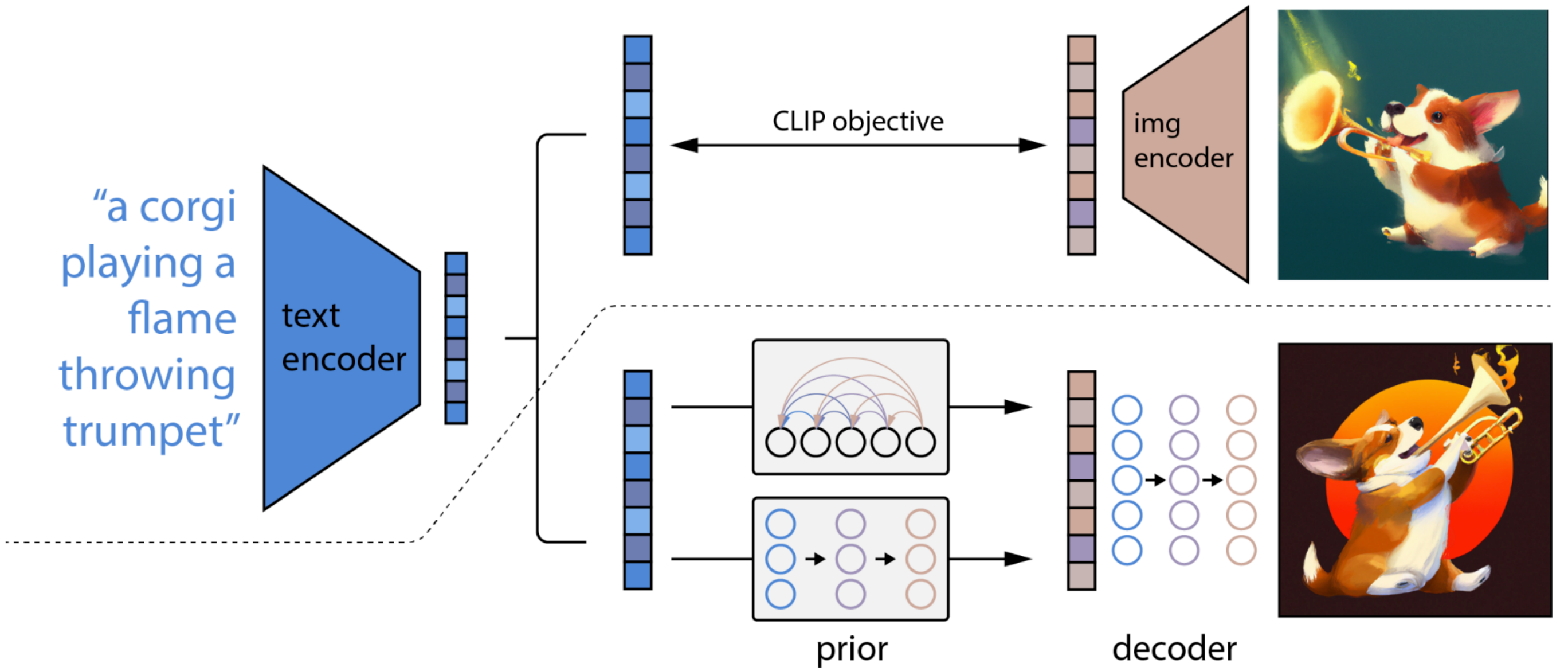}
\caption{DALLE-2 framwork. The illustration above dotted line is training process of CLIP, which is frozen during training the below prior and decoder in DALLE-2. For generation, a given text caption is passed through a CLIP text encoder. Then, the prior network generates image embedding from given text embedding. Finally, a diffusion decoder is conditioned on image embedding to generate images. (\emph{image source from: \cite{ramesh2022hierarchical}}.)}
\label{fig:dalle2}
\end{figure}

\begin{figure} 
\includegraphics[width=\linewidth]{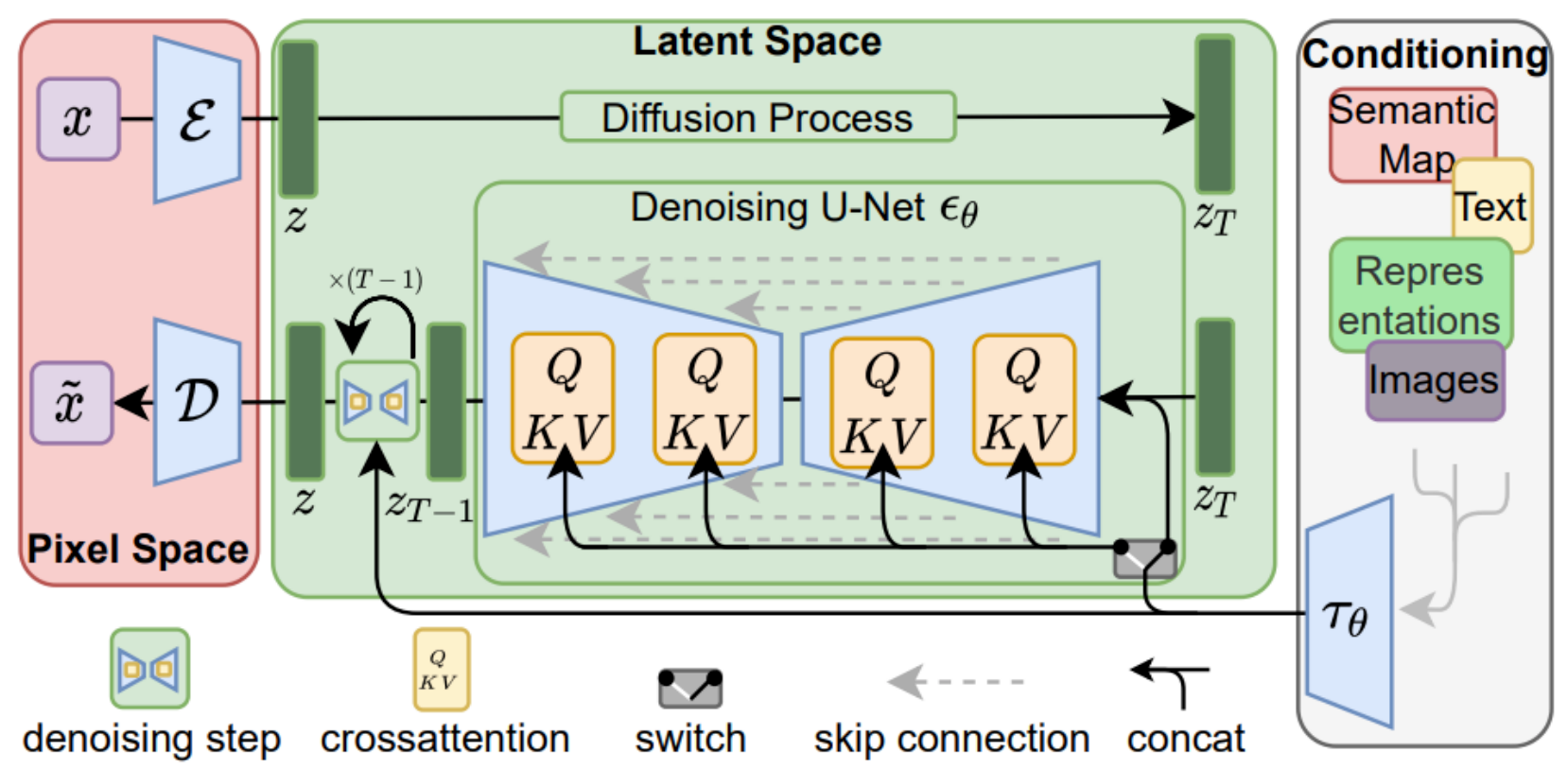}
\caption{During training, an image $x$ is fed into an encoder $\mathcal{E}$, downsampling the given $x$ into latent code $z$ with lower dimensionality. Note that $z_T$ can be efficiently obtained using similar approach in Eq.~\ref{eq:dm_x_eps}. (\emph{image source from: \cite{rombach2022high}}.)}
\label{fig:ldm}
\end{figure}

\subsubsection{Diffusion-based Approaches} \label{sec:diffusion-based}
In recent years, the use of the diffusion model (DM) for image generation has increased significantly as it is capable of generating high-quality results that outperform those produced by GANs \cite{dhariwal2021diffusion}. For text-guided generation, GLIDE \cite{nichol2021glide} trains a DM to predict $p(x_t\mid x_{t-1}, c)$, where $c$ is text caption. The text caption is first encoded as $K$ tokens and the tokens are fed into a Transformer which gives token embeddings. Then, the embeddings are projected and concatenated to each attention layer of diffusion model. Unlike GLIDE, conditioning directly on text embeddings, DALLE-2 \cite{ramesh2022hierarchical} leverages CLIP representations to achieve diverse image generation. DALLE-2 first trains a prior which generates image embeddings from text caption, and then, it trains a diffusion decoder conditioned on the image embeddings from the first step (Figure~\ref{fig:dalle2}). Though DALLE-2 produces promising, highly realistic and synthetic images, they lack certain details or demonstrate imperfections in complex scenes, resulting in a degree of artificiality. The training process of DALLE-2 may be operated on high-resolution images to mitigate this issue; however, it is very computationally intensive. Recently, \citet{rombach2022high} \footnote{Their approach becomes notable Stable diffusion.} overcomes this difficulty by introducing latent diffusion model (LDM), which performs diffusion process in latent space of lower dimensionality (Figure~\ref{fig:ldm}). The generation process from a noisy latent code can also be concatenated with different guidances, including text, images or semantic maps. 

\begin{figure} 
\includegraphics[width=\linewidth]{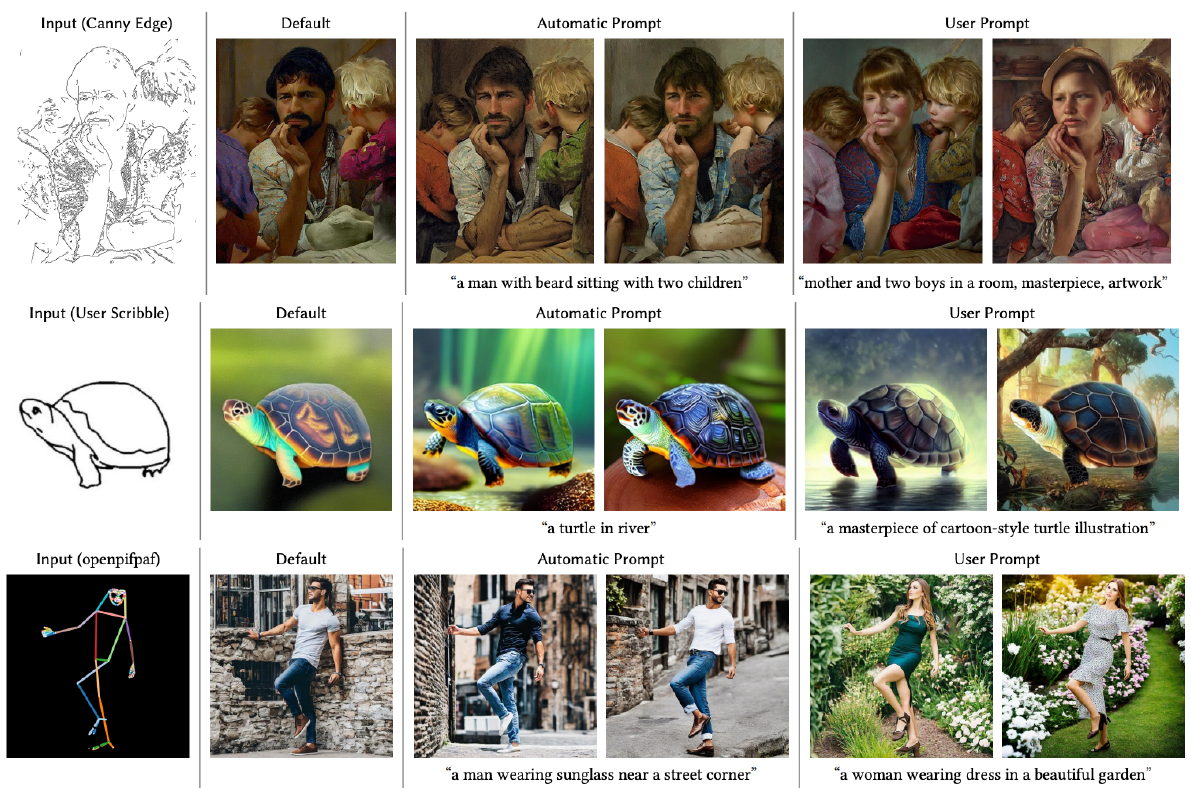}
\caption{Given a guidance, e.g. canny edge, user scribble, or poses, ControlNet is capable of controlling pre-trained Stable Diffusion for generating images under given guidance map. Text can also be concatenated with the latent code in the denoising process. Same procedure as in Figure~\ref{fig:ldm}. (\emph{image cropped from: \cite{zhang2023adding}}.)}
\label{fig:controlnet}
\end{figure}

For text-guided image editing, DiffusionCLIP \cite{kim2021diffusionclip} shares similar approach to \cite{gal2021stylegan} but operating on a pre-trained diffusion model, i.e. treating denoising network as a regularizer. The input image is first converted into a latent code via forward process (Eq.~\ref{eq:dm_forward_process}), and they propose to use CLIP loss between given and target text prompt (along with identity loss), to fine-tune the parameters in the model for the reverse process, resulting editing via target text prompt on given image. Other guidances such as sketches, \citet{voynov2022sketch} trains a small MLP for predicting edges of given images, using latent features from intermediate layers in the U-Net as inputs. For sampling, at each timestamp of denoising, the negative gradient of the loss between groundtruth edge map and the predicted edges computed from the trained MLP is added to the denoised latent code. This negative gradient guidance allows denoising process to synthesize images constrained on given edge map. For more general guidances, the state-of-the-art ControlNet \cite{zhang2023adding} employs a pre-trained diffusion model controlled by a neural network, i.e. ControlNet. ControlNet is utilized to train and compute feature offsets. These offsets are then used to concatenate with different layers in denoising U-Net to support different task-specific guidances, e.g. edges, human poses, user scribbles, etc. (Figure~\ref{fig:controlnet}). 

\begin{figure} 
\includegraphics[width=\linewidth]{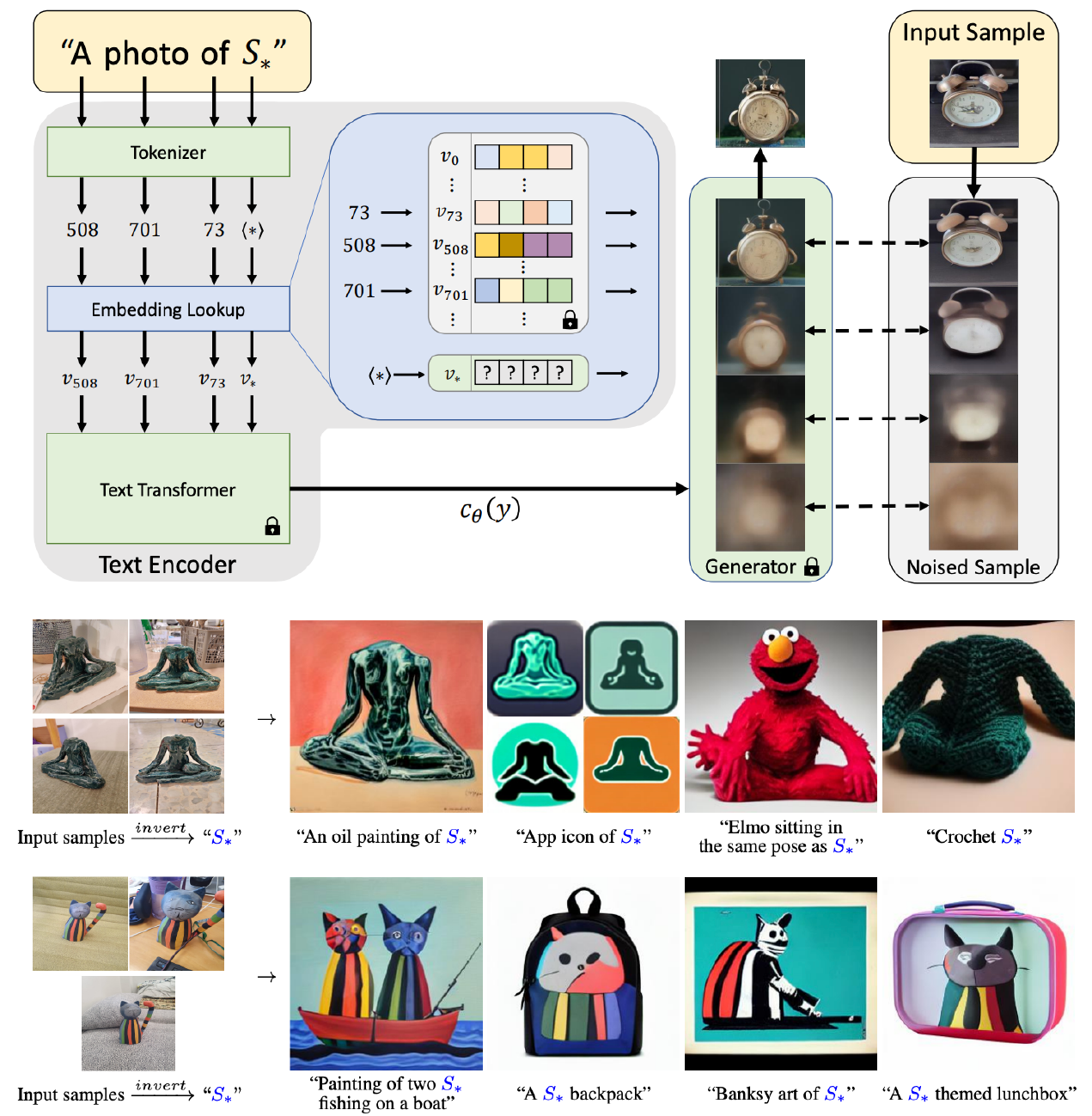}
\caption{Framework (top sub-figure) and results (bottom sub-figure) from \emph{Personalizing
Text-to-Image Generation} \cite{gal2022image}. They find the continuous representation of a text prompt involving a placeholder word "$S_*$", and pass these embeddings into a text encoder, resulting in a single conditioning code $c_\theta(y)$ which guides the denoising process in latent diffusion model. Once the embedding of the placeholder word is found, text prompt can be rephrased in various ways to personalize the generation. (\emph{image cropped from: \cite{gal2022image}}.)}
\label{fig:image_is_worth}
\end{figure}

\subsubsection{Zero-shot Approaches} \label{sec:zero-shot}
The computational demand of training and inferencing a large model has recently shown increasing impact on environment \cite{patterson2021carbon}. This is due to the fact that these computational cost require heavily on non-renewable sources, e.g. electricity from running GPUs, leading increased carbon footprint. Zero-shot approaches alleviates this issue without training any neural network to perform editing tailored to a specific task by, e.g. a pre-trained CLIP model. CLIPDraw \cite{frans2021clipdraw} optimize colors and positions of a set of random Bézier curves on a blank image to match given text description via similarity loss between embeddings of rasterized image and text prompt in CLIP space. This work motivates recent work CLIPasso \cite{vinker2022clipasso}, but with the goal of stroke-based image abstraction (Figure~\ref{fig:clipasso}). Similar to CLIPDraw, \citet{tian2022modern} optimize a set of random transparent triangles using CLIP and per-pixel $L_2$ loss to stylize images for geometric abstractions. 

Apart from using CLIP to guide parameters of interests, many recent works gain success over guiding either a generator or a latent code with CLIP loss for image editing. \citet{galatolo2021generating} use genetic algorithm to control input latent code of a pre-trained generator using CLIP loss to find images generated from GAN that matches the given text prompt. VQGAN-CLIP, proposed by \cite{crowson2022vqgan}, guides the latent code in VQGAN \cite{esser2021taming} latent space with CLIP loss to produce high-quality images from text prompt. Different from previous approaches, VQGAN-CLIP first randomly crops and augments the generated image returned from VQGAN's generator, then it computes average CLIP loss between all of the augmented images and given text prompt. To push image generation further, the ultimate goal of its editability is to synthesize realistic images from a set of \emph{personal} images. \cite{gal2022image} have shown that a pre-trained latent diffusion model (LDM) is capable of re-generating images from different text prompts given a set of 3-5 personal images. They propose to use text inversion to solve for the embedding vector of a "placeholder word", which is designed for given set of images, by minimizing the LDM loss. The placeholder word can then be used to construct different text prompts to synthesize different images while incorporating personal content (Figure~\ref{fig:image_is_worth}).

\begin{figure} 
\includegraphics[width=\linewidth]{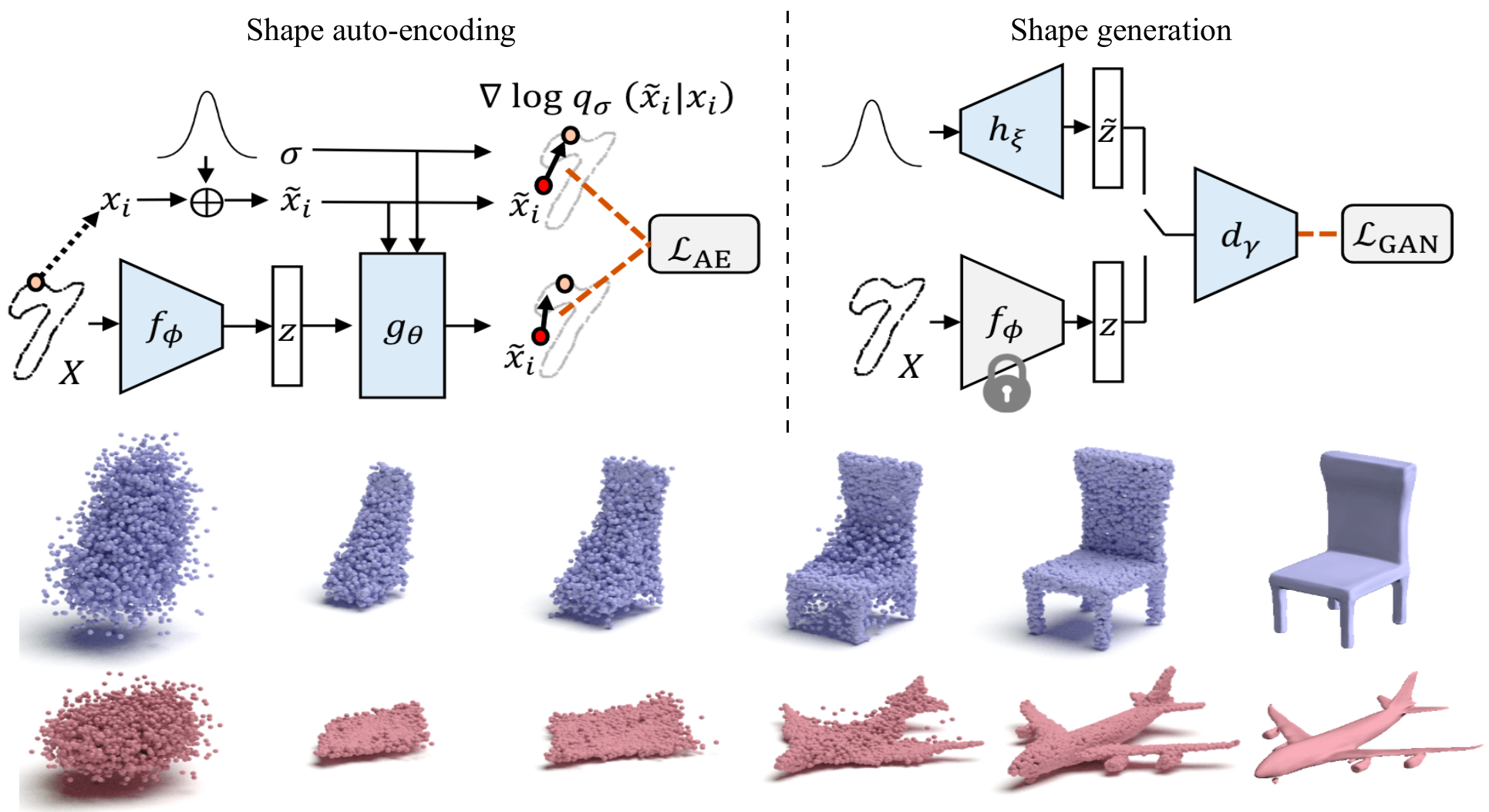}
\caption{Overall structure of \cite{cai2020learning}. During shape auto-encoding (top left), a point cloud $X$ is first encoded by an encoder $f_\phi$ into latent code $z$. Then, the gradient fields networks $g_\theta$ then take sampled points $\Tilde{x_i}$ and variance $\sigma$ as inputs to predict gradient fields with conditioning on latent code $z$. The sampled points then progressively move from random locations to shape surfaces by predicted gradients. For generation (top right), a latent-GAN is used to learn the distributions of shapes. A generator $h_\xi$ first generates synthetic latent code $\Tilde{Z}$, then sampled points $\Tilde{x_i}$ and $\sigma$ with conditioning on $\Tilde{Z}$ are then fed into gradient predictor $g_\theta$ to generate point clouds in a gradual manner (see results in the bottom row). Their approach is also compatible with generating implicit surfaces. (\emph{image cropped from: \cite{cai2020learning}}.)}
\label{fig:learning_gradients}
\end{figure}

\subsection{Shape Problems}
Text-guided image editing algorithms have achieved impressive results through learning-based approaches (Sec.~\ref{sec:image_problems}). Numerous 2D algorithms have been adapted and re-proposed for the use in 3D context. However, the lack of text-shape data pair has created difficulties in applying these methods to such scenarios. In this section, we introduce state-of-the-art approaches solving text-shape problems. These techniques include novel algorithms for generating point clouds by viewing them as samples from distributions (Sec.~\ref{sec:point-cloud}), voxel grids learned from coarse-to-fine networks (Sec.~\ref{sec:voxel}), and implicit fields such as occupancy or signed distance (Sec.~\ref{sec:implicit}). We also discuss several approaches which do not need any training pair to achieve shape editing, stylization and other applications (Sec.~\ref{sec:stylization} and ~\ref{sec:zero-shot-shape}).

\begin{figure} 
\includegraphics[width=\linewidth]{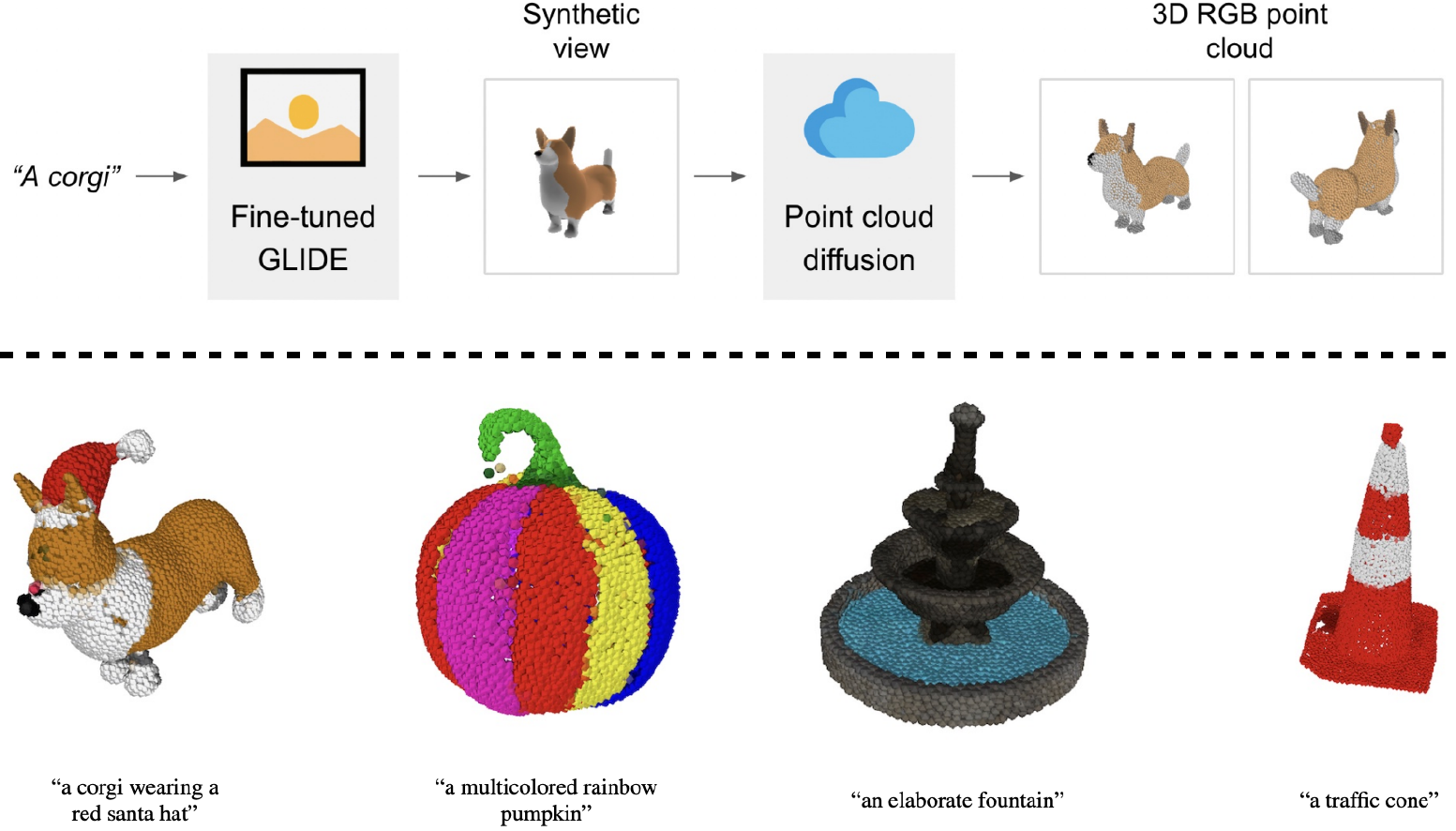}
\caption{Point$\cdot$E \cite{nichol2022point} first generates synthetic view from text captions via a fine-tuned GLIDE model \cite{nichol2021glide} (with 3 billion parameters fine-tuned on rendered 3D models from the dataset), and train a point cloud diffusion model conditioned on synthetic view to generate coarse point cloud (1024 points). Results are shown in the bottom row. (\emph{image cropped from: \cite{nichol2022point}}.)}
\label{fig:point-e}
\end{figure}

\subsubsection{Point Clouds Generation} \label{sec:point-cloud}
The application of image-based algorithms on point clouds can prove to be challenging due to their irregular structure. Pioneer work \cite{achlioptas2018learning} on generative models for point clouds propose to train an autoencoder (AE) to learn latent representation of point clouds and then train a GAN in that latent space of AE. The learned code of the trained generator is then decoded by AE's decoder to achieve point clouds reconstruction even on unseen shapes. Graph convolutional networks has also been used under GAN framework for point cloud generation \cite{shu20193d, valsesia2019learning}. While these methods show promise in reconstructing 3D point clouds, training GANs can be challenging, and permutation invariance is often not guaranteed. Some methods consider point clouds as samples from a 3D point distribution which represents the surface of a shape. Instead of solely learning latent representations of point clouds, a proposed alternative by \cite{cai2020learning} trains a gradient fields networks, $g_\theta$, to model the distribution of point clouds conditioned on the latent code of an input shape (Figure~\ref{fig:learning_gradients}). They then use a latent-GAN \cite{achlioptas2018learning} to learn the distribution of latent codes for various shapes. To generate new shapes, rather than sampling from a decoder as in \cite{achlioptas2018learning}, \citet{cai2020learning} first sample a latent code from the GAN and then feed it into the trained $g_\theta$ to sample point clouds from predicted gradient fields. Continuing with the same motivation of treating point clouds as a distribution, \citet{luo2021diffusion} formulate simple and flexible diffusion probabilistic models to generate point clouds. The forward process of their diffusion model operates on individual 3D points. To train the model on point clouds, they encode the input point clouds into a distribution using a variational auto-encoder. They then sample a latent code $z$ from this distribution, which is used to condition the network that predicts the mean of the random noise added during the forward process. For text-conditional point clouds generation, LION \cite{zeng2022lion} design a two-stage hierarchical point cloud VAE with latent diffusion models. Their approach supports conditional synthesis, shape interpolation, and mesh outputs. To facilitate practical usage, a faster approach for generating RGB point clouds from text has been proposed by \citet{nichol2022point}. Their approach involves training two diffusion models: one for generating images conditioned on text, and another for generating RGB point clouds conditioned on the generated images (Figure~\ref{fig:point-e}).

\begin{figure} 
\includegraphics[width=\linewidth]{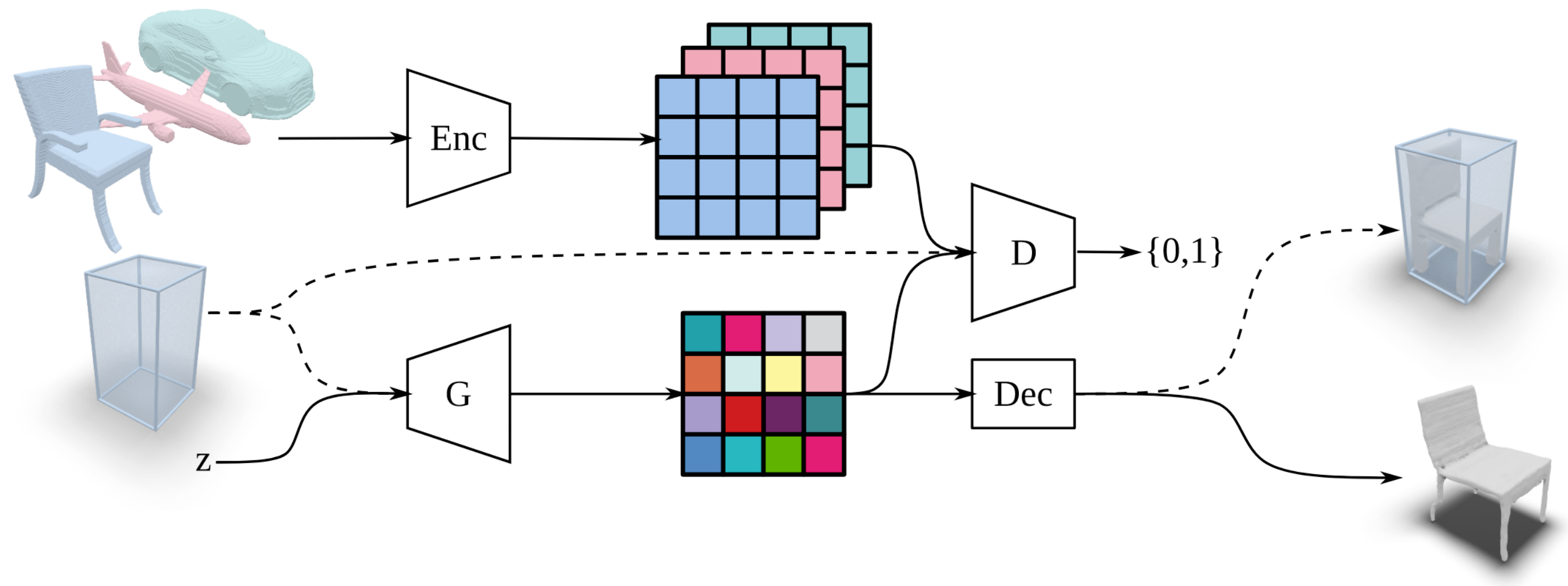}
\caption{Given high-resolution input shapes with voxel grids ($n^3$), the 'Enc' encodes input shapes into low-resolution grids of latent vectors ($k^3$, with $k \ll n$) by several strided convolutions. Then, a GAN (G and D) is trained to learn the distribution of such mappings to generate new grids of latent vectors, which is used to be fed into a decoder ('Dec') from shape generation. The generation process can be conditioned on bounding box or shape parts etc. (\emph{image source: \cite{ibing20213d}}.)}
\label{fig:voxel-grid}
\end{figure}

\subsubsection{Voxel-Grid Generation} \label{sec:voxel}
Voxel grid is 3D grid of volumetric pixels that enables efficient processing of 3D data. It is essential for tasks such as 3D reconstruction, object detection, and scene understanding. The pioneering work on 3D shape generation using voxel representation was proposed by \cite{wu2016learning}, where they generate shapes by utilizing volumetric convolutional networks under the GAN framework. To support generation from text, Text2Shape \cite{chen2019text2shape} addresses this task by learning a joint embedding space of natural language descriptions and shapes, using a dataset of text-voxel pairs. The model achieves high-quality outputs through the use of a novel conditional Wasserstein GAN framework. ShapeCrafter \cite{fu2022shapecrafter} extends Text2Shape and proposes a large text-shape dataset to provide recursive text-conditioned generation and editing through text prompts. Rather than directly generating shapes from latent space, recently, latent generative models, i.e. generating latent code, have gained popularity in generation tasks due to their ability to represent data in a compact form, avoiding the need for expensive computations. \citet{ibing20213d} propose an autoencoder framework for 3D shape generation. Their approach involves encoding shapes represented by voxel grids into low-resolution grids and training a GAN to generate grids of latent cells, which can then be decoded to generate 3D shapes under (un-)conditional settings (Figure~\ref{fig:voxel-grid}). In line with the objective of coarse-to-fine generation, \citet{sanghi2022textcraft} propose a three-stage training approach. First, two separate VQ-VAEs \cite{van2017neural} are trained for low-resolution and high-resolution voxel inputs, respectively. Second, a coarse transformer, which is conditioned on a CLIP embedding, is trained to generate low-resolution VQ-VAE latent grids. Finally, a fine transformer, which is conditioned on the latent grids from the second step, is trained to generate high-resolution VQ-VAE latent grids. Their hierarchical latent generative models have produced outputs with stable quality and increased diversity in generation compared to CLIP-Forge \cite{sanghi2022clip}.

\begin{figure} 
\includegraphics[width=\linewidth]{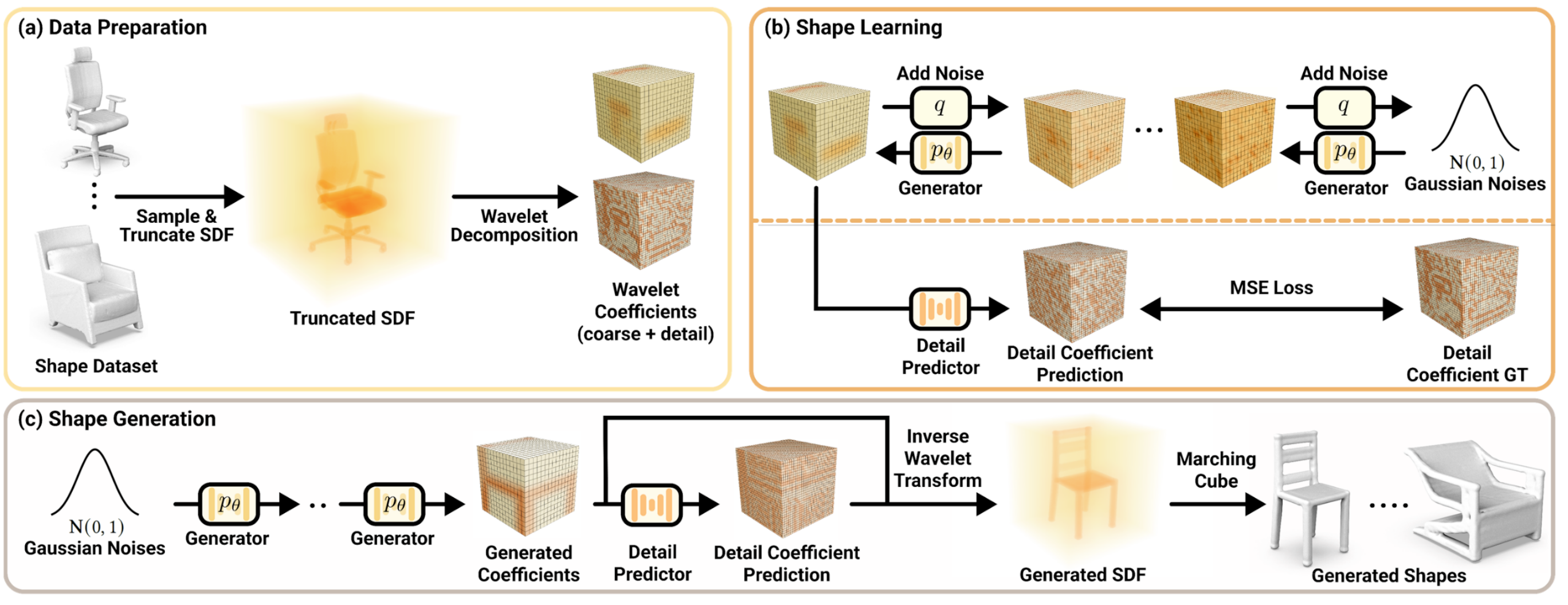}
\caption{Training process of \citet{hui2022neural}. Rather than following typical methods of training autoencoders, \citet{hui2022neural} utilize wavelet decomposition to compute both coarse and fine wavelet coefficients of a truncated SDF. Then, a diffusion model and a detail predictor are trained to predict coarse wavelet coefficients from random noise and fine wavelet coefficients, respectively. Generation process follows the previous two steps intuitively. (\emph{image source: \cite{hui2022neural}}.)}
\label{fig:wavelet}
\end{figure}

\subsubsection{Implicit-representation Generation} \label{sec:implicit}
\begin{figure} 
\includegraphics[width=\linewidth]{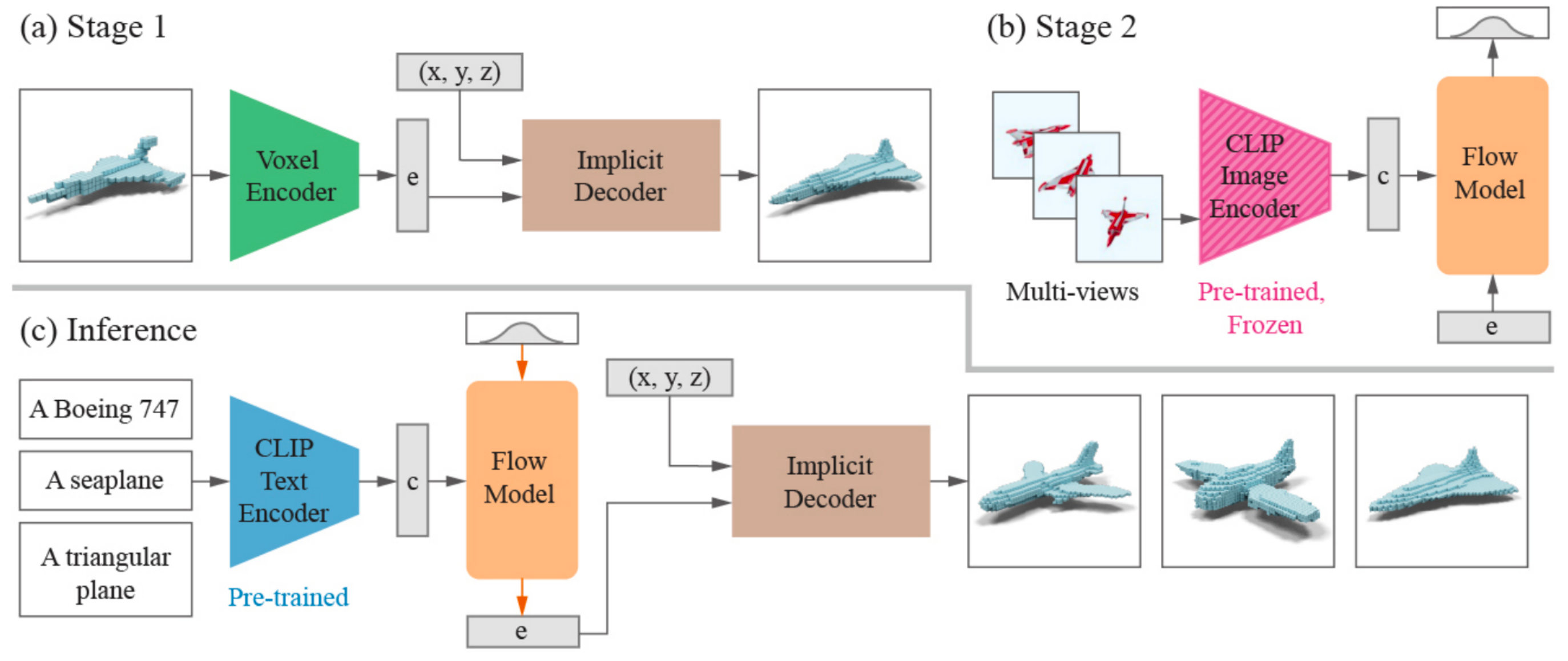}
\caption{CLIP-Forge \cite{sanghi2022clip} framework. The top diagram shows the two-stage training process. The first training stage is to encode the given voxel inputs and decode the implicit latent embeddings $e$ via a implicit decoder. Second stage is to train a flow model from $e$ to normal distribution conditioned on CLIP embeddings $c$. Inference step (shown in bottom diagram) simply combines the above two stages. (\emph{image source: \cite{sanghi2022clip}}.)}
\label{fig:clip-forge}
\end{figure}

Implicit functions are a popular choice for representing 3D shapes because of their ability to model smooth surfaces even for complex shapes. There are two common types of implicit functions used for representing shapes: signed distance fields (SDF), which assign each point in the fields a scalar value indicating the closest distance to the shape surface (positive for outside and negative for inside), and occupancy fields, which assign each point either 0 (inside the shape) or 1 (outside the shape). Both representations can be reconstructed into surfaces using (neural) marching cube \cite{lorensen1987marching, chen2021neural} or (neural) dual contouring \cite{ju2002dual, chen2022neural}. \citet{hui2022neural} share similar coarse-to-fine approach to \cite{sanghi2022textcraft} but operate on wavelet domain using diffusion model (Figure~\ref{fig:wavelet}). For text-guided generation, CLIP-Forge \cite{sanghi2022clip} encodes input voxel grids into implicit shape embeddings and trains a flow model (conditioned on CLIP embeddings) that turns the implicit embeddings into normal distribution. This two-stage training framework finds implicit shape embeddings only through unlabelled data, providing diverse zero-shot generation using pre-trained CLIP (Figure~\ref{fig:clip-forge}). \citet{liu2022towards} propose a learning framework to decouple shape and color features along with a word-level spatial transformer to link these features, enabling text-guided generation with colors and capability for editing. Recent success of neural radiance fields (NeRF) \cite{mildenhall2021nerf} yields highly realistic 3D reconstructions from a limited numbers of 2D image views using neural networks. \citet{jain2022zero} propose Dream Fields, which incorporates CLIP guidance into the NeRF framework. This technique aims to reduce both transmittance loss and the discrepancy between multi-view image embeddings and text embeddings, ultimately resulting in the generation of text-guided NeRF models.

\begin{figure} 
\includegraphics[width=\linewidth]{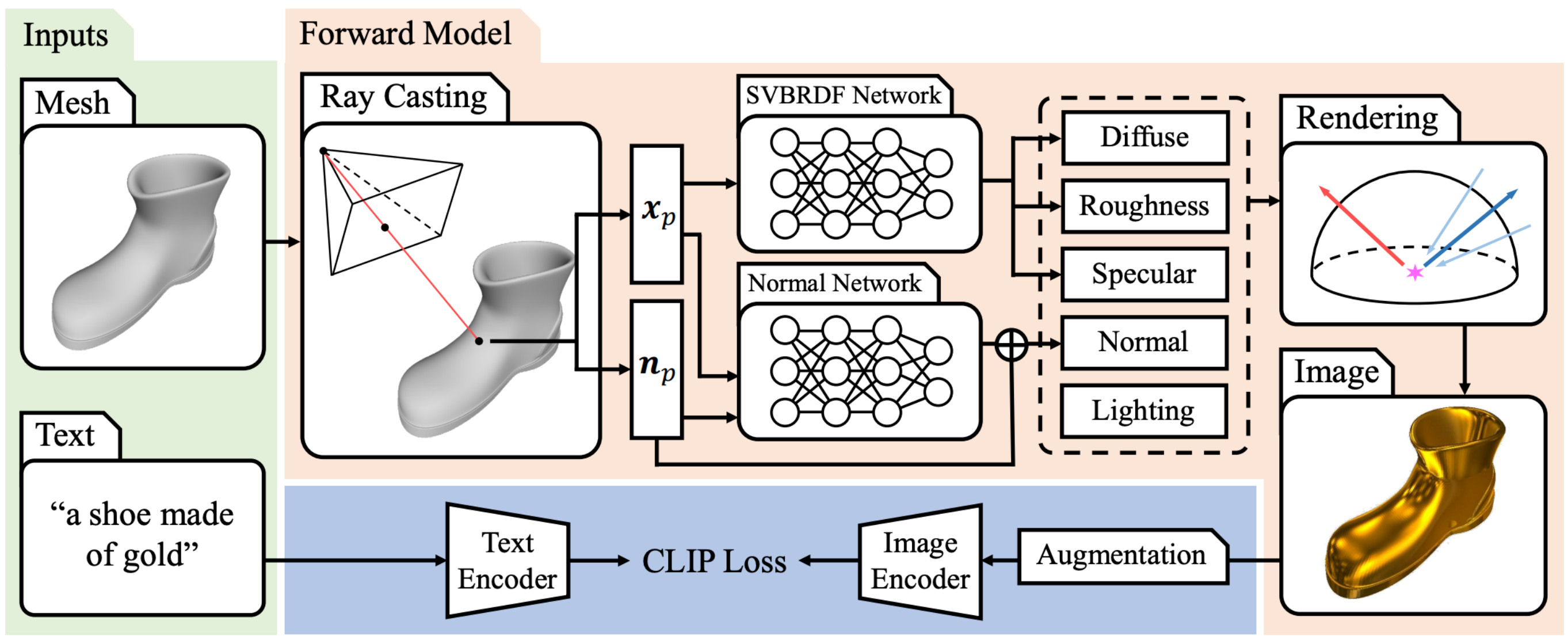}
\caption{Given a mesh and a text prompt, TANGO first casts rays to the intersection point $x_p$ on the mesh with normal $n_p$. Then, $x_p$ and $n_p$ are fed into two MLPs to predict SVBRDF parameters as well as normal variations. The rendering equation is approximated by spherical Gaussian \cite{zhang2021physg}. Since the whole framework is differentiable, the gradients are backpropagated from the loss computed in CLIP embedding space between text and rendered images. (\emph{image source: \cite{chen2022tango}}.)}
\label{fig:tango}
\end{figure}

\subsubsection{Stylization} \label{sec:stylization}
The task of stylizing shapes based on text prompts is made difficult by the lack of clean text-shape data pairs. Recently, Text2Mesh \cite{michel2022text2mesh} has overcome this challenge by optimizing a neural style network with a CLIP-based semantic loss. This approach does not require pre-trained generative models and can stylize a given mesh according to a given text prompt. Similar motivation to Text2Mesh, TEXTure \cite{richardson2023texture} achieves higher quality texturing 3D shapes from a given text prompt by employing a pre-trained depth-to-image diffusion model. They paint the given mesh iteratively from different camera views with the image generated by a depth-conditioned diffusion model along with text, normal and trimap inputs. Rather than focusing on geometric details of meshes, TANGO, proposed by \cite{chen2022tango}, predicts spatially varying bidirectional reflectance distribution function (SVBRDF) parameters and normal variations by two MLPs to render images from different views using differentiable renderer. Then, the rendered images are encoded through CLIP image encoder to compute loss with given text embedding (Figure~\ref{fig:tango}). Similarly, CLIP-Mesh \cite{mohammad2022clip} optimizes the texture, normal, and vertex positions of a given mesh from text prompt under a laplacian regularizer. Their approach is achieved through using a differentiable renderer that generates multiple-view images, which are then embedded into CLIP space. The resulting image embeddings are used to compute the loss with a text embedding and a diffusion prior.

\subsubsection{Other Applications} \label{sec:zero-shot-shape}
Parameters of interest are the key design elements for utilizing CLIP to supervise various subjects of text-guided generation. AvatarCLIP \cite{hong2022avatarclip} presents a novel method for creating 3D animations of avatars using text as input. Their approach optimizes a shape VAE to generate a coarse avatar shape according to given text with shape attributes. Similar to many prior works in Sec.~\ref{sec:stylization}, the avatar shape and texture are then sculpted using CLIP with multi-view images but in the implicit space. Finally, a CLIP-guided motion synthesis framework is proposed to generate avatar animation. Another related work, MotionCLIP \cite{tevet2022motionclip}, generates human motions from text by training a transformer-based autoencoder where its latent space is trained to align with the embedding space of CLIP. 

\begin{figure} 
\includegraphics[width=\linewidth]{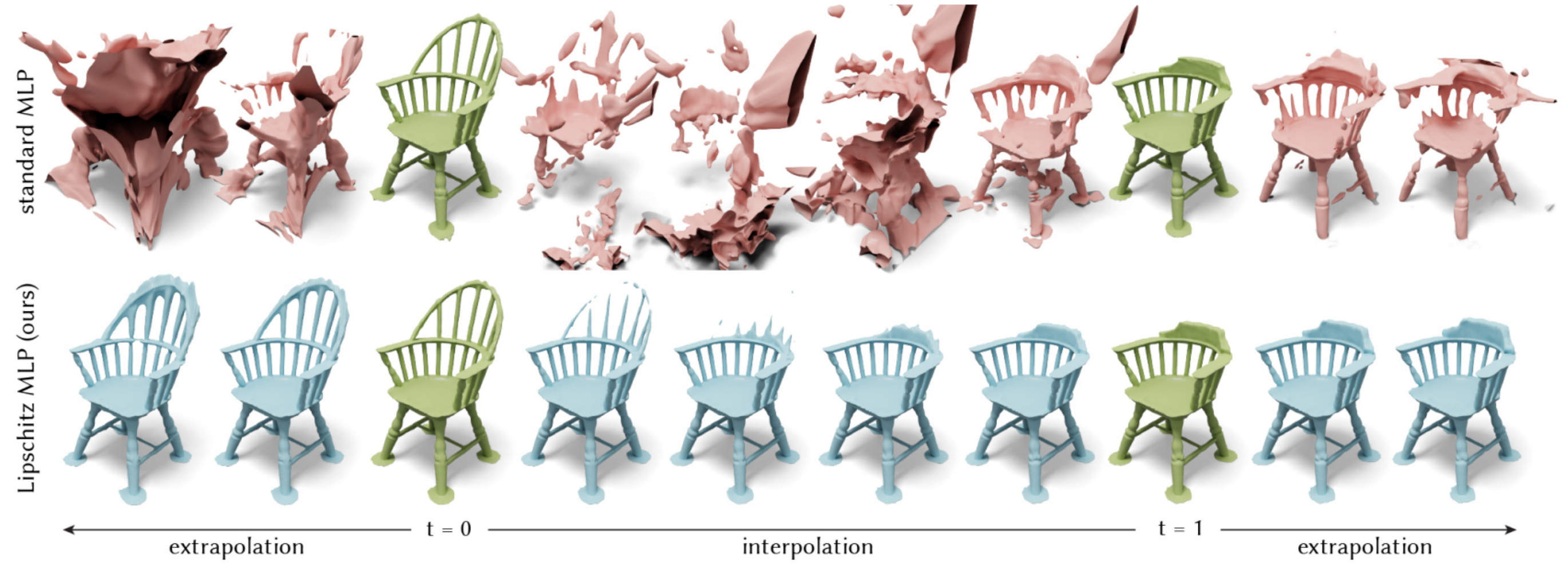}
\caption{An MLP takes input of coordinates along with latent code $t$ and outputs signed distance fields. In this diagram, the MLP is trained on two shapes, where $t=0$ is the higher chair and $t=1$ is the lower one. \citet{liu2022learning} apply Lipschitz bound on latent code $t$ to encourage smooth transition between shapes (bottom row) while standard MLP (upper row) fails to interpolate meaningful structure. (\emph{image source: \cite{liu2022learning}}.)}
\label{fig:lipschitz}
\end{figure}

\begin{figure} 
\includegraphics[width=\linewidth]{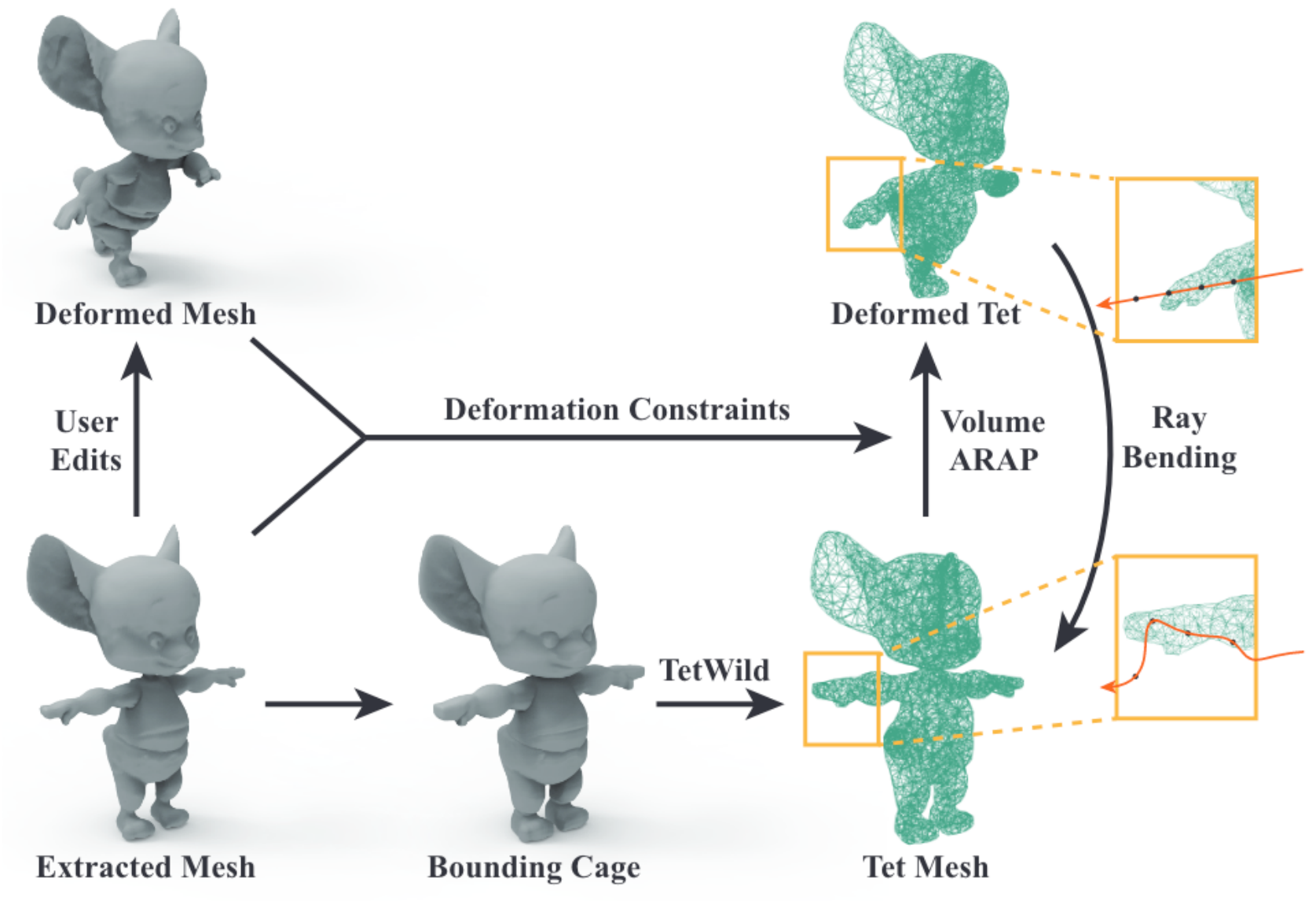}
\caption{Framework of NeRF-Editing \cite{yuan2022nerf}. Given a trained NeRF, the mesh is extracted from the zero-level set of SDF computed from NeuS \cite{wang2021neus}. Users are allowed to deform the mesh under ARAP formulation. When performing ray marching, the barycentric coordinates of each sample point are computed with respect to the deformed tetrahedron which it lies in. The computed barycentric coordinates then are used to apply on the displacement vectors between original tetrahedron vertices and deformed tetrahedron vertices to obtain an updated camera view, i.e. ray bending. (\emph{image source: \cite{yuan2022nerf}}.)}
\label{fig:nerf-editing}
\end{figure}

\section{Editing under INR}
Implicit neural representations (INR) is a specific type of representations that use neural networks, typically an MLP, to encode various types of data such as image values \cite{dupont2021coin, sitzmann2020implicit}, volumetric rendering parameters (e.g. NeRF) \cite{mildenhall2021nerf}, signed distances \cite{park2019deepsdf}, or occupancy \cite{mescheder2019occupancy}. These representations exhibit high memory efficiency and facilitate end-to-end optimization due to the differentiability of the MLP. Nonetheless, direct editing in the context of INR remains challenging, as the weights in a trained MLP generally hold little significance for editing purposes. Several recent works make progress on editing under INR. \citet{liu2022learning} use Lipschitz regularization on the latent code while training an MLP, enabling smooth interpolation between shapes (Figure~\ref{fig:lipschitz}). \citet{jung2022deep} propose a learning framework for deforming 3D caricatures. They train a hypernetwork \cite{ha2016hypernetworks} that outputs the parameters of SIREN MLP \cite{sitzmann2020implicit}. The SIREN MLP then takes the coordinate of caricature's meshes and predicts the offset vector, which is used to modify and create the deformed caricature. In the subsequent subsection, we commence by providing an overview of NeRF and then present an introduction to the cutting-edge research related to NeRF editing.
\paragraph{Volumetric rendering} 
NeRF is a neural network trained on a limited numbers of images from different viewing directions to encode parameters for volumetric rendering. NeRF takes a coordinate $x_i$ as well as camera view $d$ as inputs, and return view-dependent color $c(x_i,d)$ and volume density $\sigma(x_i)$. To render a scene from a viewing direction $d$, we perform ray marching from a ray $r(o,d) = o + t_i \cdot d$. At each sampled point $x_i = o + t_i \cdot d$ along the ray (assume there are $N$ samples), we query the trained NeRF to obtain $c(x_i,d)$ and $\sigma(x_i)$. The reconstructed color $\mathbf{c}$ from that ray $r$ can then be computed using volume rendering equation (\cite{kuang2022palettenerf}):
\begin{equation} \label{eq:volume_rendering}
\begin{aligned}
\mathbf{c}(r) \approx \sum_{i=1}^N \alpha_i\cdot(1-w_i)\cdot c(x_i, d)
\end{aligned}
\end{equation}
where $w_i = \exp(-(t_i-t_{i-1})\cdot \sigma(x_i))$ is the transmittance of the ray between the $i^{\text{th}}$ and $(i-1)^{\text{th}}$ samples and $\alpha_i = \Pi_{j=1}^{i-1}w_j$ represents the attenuation of the ray from its origin to the $i^{\text{th}}$ sample. To train a NeRF, we can simply minimize the loss between reconstructed colors computed from Eq.~\ref{eq:volume_rendering} and the ground truth colors.

In the context of NeRF editing, \citet{yuan2022nerf} develop a method that enables editing of NeRF without the need for re-training. Their approach involves extracting meshes from the NeRF and allowing users to edit the explicit mesh using ARAP deformation method \cite{sorkine2007rigid}. Then, ray bending is performed to estimate the volumetirc rendering parameters on the deformed mesh (Figure~\ref{fig:nerf-editing}). Recent NeRFshop \cite{jambon2023nerfshop} performs similar approach to \cite{yuan2022nerf} but allows fast interactive editing under framework of instant NGP \cite{muller2022instant}. The work uses pre-computed mean value coordinates as deformed handles and an adaptive lookup table for tetrahedron intersection to speed up the sample query. Instead of directly manipulating NeRF, numerous recent studies \cite{wu2022palettenerf, gong2023recolornerf} have focused on addressing the recoloring problem on NeRF. Specifically,  PaletteNeRF \cite{kuang2022palettenerf} decompose the color function $c(x, d)$ in Eq.~\ref{eq:volume_rendering} into a trainable formulation of view-dependent colors with weights and palette colors, allowing real-time NeRF color editing via palettes.

\section{Discussion and Conclusion}
We have shown various state-of-the-art methods for both editing images and shapes. Though multiple approaches have been suggested for tackling image editing challenges, they are all driven by a shared objective of identifying a way for optimizing the latent representation of an image through targeted regularization, with the aim of accomplishing specific tasks. Recent works have been more focused on personalized image editing, e.g. \cite{sarukkai2023collage, iluz2023word}, using large pre-trained models. While these models have proven to have a highly meaningful latent space, they are extremely data-hungry, and there are still numerous challenges to overcome in controlling the behaviors of the underlying generating process. Several recent works have progressed on understanding these large models through human interaction \cite{lee2023aligning} or re-train it with augmented framework \cite{zhang2023adding}. 

When it comes to shape-related problems, current research largely focuses on identifying effective, compact representations that can be leveraged for both generation and editing purposes. However, the lack of text-shape data has made it difficult to train a latent space in which embeddings are properly aligned with natural languages. Though prior works such as \cite{tevet2022motionclip} have shown promise in mitigating this issue through the use of CLIP, the challenge remains since rendered images from different viewing directions of a shape might not hold similar embeddings in CLIP space. 

Another promising direction for future research is to explore the possibility of directly learning on data under INR. While INR has been demonstrated to be highly effective at compressing images and scenes through the use of neural networks, there are still significant challenges to overcome such as data augmentations on INR. Unlike images, which have pixels that can be naturally operated with convolutions, INR currently lacks a canonical input format that can be used for model learning. Recent work \cite{de2023deep} proposes to directly encode MLP's weights along with bias into a latent code and train a implicit decoder with coordinates to reconstruct the implicit function. This represents a promising breakthrough in the field, and it is likely that further research will build on this approach in the years to come.

This short survey offers a comprehensive high-level overview of the state-of-the-art editing and generation techniques as presented in over 50 papers. We have divided the survey into two main categories: 2D and 3D problems. Specifically, image problems have been categorized based on the different models used, while shape problems are organized according to their respective representations. Many directions remain promising such as personalized editing, more targeting shape generation, and NeRF editing. As these fields continue to evolve and develop, we look forward to the new breakthroughs and insights that will help advance our understanding on editing and generation from a human cognition perspective.
\bibliographystyle{ACM-Reference-Format}
\bibliography{bib/merged}

\end{document}